\documentclass[useAMS,usenatbib]{mn2e}
 
\usepackage{times} 
\usepackage{epsf}
 
\def\etal{{\it et al.\ }}
\def\gsim{\ga}
\def\lsim{\la}
\def\Msol{M_\odot}
\def\Zsol{Z_\odot}
\def\ergs{{\rm erg\,s^{-1}}}
\def\ergcms{{\rm erg\,cm^{-2}s^{-1}}}
\def\Hz{{\rm Hz}}
\def\kpc{{\rm kpc}}
\def\Mpc{{\rm Mpc}}
\def\kms{{\rm km\,s^{-1}}}
\def\yr {{\rm yr}}

\title[The properties of Ly$\alpha$-emitting galaxies in 
hierarchical galaxy formation models]
{The properties of Ly$\alpha$ emitting galaxies in 
hierarchical galaxy formation models}

\author[Le Delliou  et al.]{
M. Le Delliou\thanks{E-mail: delliou@alfmail.cii.fc.ul.pt (MLeD)},
C. G. Lacey,
C. M. Baugh
and S. L. Morris\\
Institute for Computational Cosmology, Department of Physics,
University of Durham, South Road, Durham, DH1 3LE, UK. 
}

\begin{document}

\maketitle
  
\begin{abstract}
We present detailed predictions for the properties of
Ly$\alpha$-emitting galaxies in the framework of the $\Lambda$CDM
cosmology, calculated using the semi-analytical galaxy formation model
{\tt GALFORM}. We explore a model which assumes a top-heavy IMF
in starbursts, and which has previously been shown to explain the
sub-mm number counts and the luminosity function of Lyman-break
galaxies at high redshift.  We show that this model, with the
simple assumption that a fixed fraction of Ly$\alpha$ photons escape
from each galaxy, is remarkably successful at explaining the observed
luminosity function of Ly$\alpha$ emitters over the redshift range
$3<z<6.6$. We also examine the distribution of Ly$\alpha$
equivalent widths and the broad-band continuum magnitudes of emitters,
which are in good agreement with the available observations. We look
more deeply into the nature of Ly$\alpha$ emitters, presenting
predictions for fundamental properties such as the stellar mass 
and radius of the emitting galaxy and the mass of the host dark
matter halo. The model predicts that the clustering of Ly$\alpha$
emitters at high redshifts should be strongly biased relative to the
dark matter, in agreement with observational estimates. We also
present predictions for the luminosity function of Ly$\alpha$ emitters
at $z>7$, a redshift range which is starting to be be probed by
near-IR surveys and using new instruments such as DAzLE.
\end{abstract}


\section{Introduction }

After an unpromising start, searches for Ly$\alpha$ emission are now
proving to be a powerful means of detecting star-forming galaxies at
high redshift \citep[e.g.][]{Hu98, Pascarelle98, Kudritzki00},
competing in observing efficiency with techniques such as broad-band
searches for Lyman-break galaxies.  The next generation of
near-infrared instrumentation \citep[e.g.][]{DAZLE} will in principle
allow Ly$\alpha$ emitting galaxies to be found up to $z \sim 20$,
permitting a probe of the star formation history of the Universe
before the epoch when reionization is thought to have taken place.

There are in fact a number of different mechanisms which can produce
Ly$\alpha$ emission from high redshift objects. (1) Gas in galaxies
which is photo-ionized by young stars will emit Ly$\alpha$ as
hydrogen atoms recombine; this was originally proposed as a signature
of primeval galaxies by \citet{Partridge67}. (2) Gas can alternatively
be ionized by radiation from an active galactic nucleus (AGN). 
(3) Intergalactic gas clouds are predicted to emit Ly$\alpha$
recombination radiation due to ionization of the gas by the
intergalactic ultraviolet background \citep[e.g.][]{Hogan87,
  Cantalupo05}. (4) Gas within a dark matter halo which is cooling and
collapsing to form a galaxy may radiate much of the gravitational
collapse energy by collisionally-excited Ly$\alpha$ emission
\citep[e.g.][]{Haiman00, Fardal01}. (5) Finally, Ly$\alpha$ can also
be emitted from gas which has been shock heated by galactic winds or
by jets in radio galaxies \citep[e.g.][]{McCarthy87}. The majority of
high-redshift Ly$\alpha$ emitters (LAEs) detected so far are compact,
and appear to be individual galaxies in which the Ly$\alpha$ emission
is powered by photoionization of gas by young stars. Ly$\alpha$
surveys have also found another class of emitter, the so-called
Ly$\alpha$ blobs, in which the Ly$\alpha$ emission is much more
extended than individual galaxies, and may be powered partly by AGNs
or gas cooling \citep{Steidel00, Bower04, Matsuda04}. We will be
focusing in this paper on Ly$\alpha$ emission powered by young stars,
and so will not consider the Ly$\alpha$ blobs further.

To date, there has been relatively little theoretical work on trying
to predict the properties of star-forming Ly$\alpha$-emitting
galaxies within a realistic galaxy formation framework.
\citet{Haiman99} made predictions for the number of emitters based on
the halo mass function and using ad-hoc assumptions linking Ly$\alpha$
emission to halo mass, while \citet{Barton04} made predictions for
very high redshifts ($z>7$) based on a gas-dynamical simulation.
\citet{Furlanetto05} used gas-dynamical simulations to calculate
Ly$\alpha$ emission both from star-forming objects and from the
intergalactic medium in the redshift range $0<z<5$.  However, the
first calculation of the abundance of Ly$\alpha$ emitters based on a
detailed hierarchical galaxy formation model was that of
\citet[][hereafter Paper I]{LeD05}.  In Paper~I, we used the {\tt
GALFORM} semi-analytical galaxy formation model to predict the abundance
of star-forming Ly$\alpha$ emitters as a function of redshift in the
cold dark matter (CDM) model. The {\tt GALFORM} model computes the
assembly of dark matter halos by mergers, and the growth of galaxies
both by cooling of gas in halos and by galaxy mergers. It calculates
the star formation history of each galaxy, including both quiescent
star formation in galaxy disks and also bursts triggered by galaxy mergers,
as well as the feedback effects of galactic winds driven by supernova
explosions. In Paper~I, we found that a very simple model, in which a
fixed fraction of Ly$\alpha$ photons escape from each galaxy,
regardless of its other properties, gave a surprisingly good match to
the total numbers of Ly$\alpha$ emitters detected in different surveys
over a range of redshifts.  We also explored the impact of varying
certain parameters in the model, such as the redshift of reionization
of the intergalactic medium, on the abundance of emitters.

In this paper, we explore in more detail the fiducial model of Paper~I
(based on an $\Omega_{m}=0.3$, spatially flat, $\Lambda$CDM model with
a reionization redshift of $10$). We use the full capability of the
{\tt GALFORM} model to predict a wide range of galaxy properties,
connecting various observables to Ly$\alpha$ emission. The galaxy
formation model we use is the same as that proposed by
\citet{Baugh05}. A critical assumption of this model is that
stars formed in starbursts have a top-heavy initial mass function
(IMF), while stars formed quiescently in galactic disks have a solar
neighbourhood IMF. We showed in \citeauthor{Baugh05} that, within the
framework of $\Lambda$CDM, the top-heavy IMF is essential for matching
the counts and redshifts of sub-millimetre galaxies and the luminosity
function of Lyman break galaxies at $z=3$ (once dust extinction is
included), while remaining consistent with galaxy properties in the
local universe such as the optical and far-IR luminosity functions and
galaxy gas fractions and metallicities. More detailed comparisons of
this model with observations of Lyman-break galaxies and of galaxy
evolution in the IR will be presented in Lacey \etal (2005a, 2005b, in
preparation).  The assumption of a top-heavy IMF is controversial,
but underpins the success of the model in explaining the 
high-redshift sub-mm and Lyman-break galaxies. It is therefore
important to test this model against as many observables as
possible. \citet{Nagashima05a} showed that a top-heavy IMF seems to be
required to explain the metal content of the hot intracluster gas in
galaxy clusters, and \citet{Nagashima05b} showed that a similar
top-heavy IMF also seems to be necessary to explain the observed
abundances of $\alpha$-elements (such as Mg) in the stellar
populations of elliptical galaxies.   In the present paper, we
explore the predictions of the \citet{Baugh05} model for the
properties of Ly$\alpha$-emitting galaxies and compare them with
observational data. We emphasize that our aim here is to explore in
detail a particular galaxy formation model which has been shown to
satisfy a wide range of other observational constraints, rather than
to conduct a survey of Ly$\alpha$ predictions for different model
parameters.

In Section~\ref{sec:GALFORM}, we give an outline of the {\tt GALFORM}
model, focusing on how the predictions we present later on are
calculated. Section~\ref{sec:lf-evoln} examines the evolution of the
Ly$\alpha$ luminosity function, and compares the model predictions
with observational data over the redshift range $3 \lsim z \lsim
7$. In Section~\ref{sec:obs-props}, we compare a selection of observed
properties of Ly$\alpha$ emitters with the model predictions. In
Section~\ref{sec:phys-props}, we look at some other predictions of the
model, most of which  cannot currently be compared directly with
observations. Section~\ref{sec:high-z} extends the predictions for the
Ly$\alpha$ luminosity function to $z> 7$. We present our
conclusions in Section~\ref{sec:conc} .

\section{Galaxy formation model}
\label{sec:GALFORM}

We use the semi-analytical model of galaxy formation, {\tt GALFORM},
to predict the Ly$\alpha$ emission and many other properties of
galaxies as a function of redshift.  The general methodology and
approximations behind the {\tt GALFORM} model are set out in detail in
\citet{Cole00}. The particular model that we use in this paper is the
same as that described by \citet{Baugh05}.  The background cosmology
is a cold dark matter universe with a cosmological constant
($\Omega_{m}=0.3$, $\Omega_{\Lambda}=0.7$, $\Omega_{b}=0.04$, $h
\equiv H_0/100\kms\Mpc^{-1}=0.7$, $\sigma_8=0.93$).
Below we review the physics behind the particular model predictions
that we highlight in this paper.

The {\tt GALFORM} model follows the main processes which shape the
formation and evolution of galaxies. These include: (i) the collapse
and merging of dark matter halos; (ii) the shock-heating and radiative
cooling of gas inside dark halos, leading to formation of galaxy
disks; (iii) quiescent star formation in galaxy disks; (iv) feedback
both from supernova explosions and from photo-ionization of the IGM; (v)
chemical enrichment of the stars and gas; (vi) galaxy mergers driven
by dynamical friction within common dark matter halos, leading to
formation of stellar spheroids, and also triggering bursts of star
formation.  The end product of the calculations is a prediction of the
number of galaxies that reside within dark matter haloes of different
masses. The model predicts the stellar and cold gas masses of the
galaxies, along with their star formation and merger histories, and
their sizes and metallicities.

The prescriptions and parameters for the different processes which we
use in this paper are identical to \citet{Baugh05}.  Feedback is
treated in a similar way to \citet{Benson03}: energy injection by
supernovae reheats some of the gas in galaxies and returns it to the
halo, but also ejects some gas from halos as a ``superwind'' - the
latter is essential for reproducing the observed cutoff at the bright
end of the present-day galaxy luminosity function. We also include
feedback from photo-ionization of the IGM: following reionization
(i.e. for $z<z_{\rm reion}$), we assume that gas cooling in halos with
circular velocities $V_c < 60\, \kms$ is completely suppressed. We
assume in this paper that reionization occurs at $z_{\rm reion}=10$,
chosen to be intermediate between the low value $z\sim 6$ suggested by
measurements of the Gunn-Peterson trough in quasars \citep{Becker01}
and the high value $z\sim 20$ suggested by the WMAP measurement of
polarization of the microwave background \citep{Kogut03}. Our model
has two different IMFs: quiescent star formation in galactic disks is
assumed to produce stars with a solar neighbourhood IMF (we use the
\citet{Kennicutt83} paramerization, with slope $x=0.4$ below $1
M_{\odot}$ and $x=1.5$ above), whereas bursts of star formation
triggered by galaxy mergers are assumed to form stars with a
top-heavy, flat IMF with slope $x=0$ (where the Salpeter slope is
$x=1.35$).  In either case, the IMF covers the mass range $0.15 <m<
120 \Msol$. As mentioned in the Introduction, the choice of a flat IMF
in bursts is essential for the model to reproduce the observed counts
of galaxies at sub-mm wavelengths. The parameters for star formation
in disks and for triggering bursts and morphological transformations
in galaxy mergers are given in \citet{Baugh05}.

The sizes of galaxies are computed as in \citet{Cole00}: gas which
cools in a halo is assumed to conserve its angular momentum as it
collapses, forming a rotationally-supported galaxy disk; the radius of
this disk is then calculated from its angular momentum, including the
gravity of the disk, spheroid (if any) and dark halo. Galaxy spheroids
are built up both from pre-existing stars in galaxy mergers, and from
the stars formed in bursts triggered by these mergers; the radii of
spheroids formed in mergers are computed using an energy conservation
argument. In calculating the sizes of disks and spheroids, we include
the adiabatic contraction of the dark halo due to the gravity of the
baryonic components.

Given the star formation and metal enrichment history of a galaxy,
{\tt GALFORM} computes the spectrum of the integrated stellar
population using a population synthesis model based on the Padova
stellar evolution tracks \citep[see][for
details]{Granato00}. Broad-band magnitudes are then computed by
redshifting the galaxy spectrum and convolving it with the filter
response functions. We include extinction of the stellar continuum by
dust in the galaxy; this is computed based on a two-phase model of the
dust distribution, in which stars are born inside giant molecular
clouds and then leak out into a diffuse dust medium \citep[see][for
more details]{Granato00}. The optical depth for dust extinction of the
diffuse component is calculated from the mass and metallicity of the
cold gas and the sizes of the disk and bulge. We note that the
extinction predicted by our model in which the stars and dust are
mixed together is very different from what one obtains if all of the
dust is in a foreground screen (as is commonly assumed in other
theoretical models). Finally, we also include the effects on the
observed stellar continuum of absorption and scattering of radiation
by intervening neutral hydrogen along the line of sight to the galaxy;
we calculate this IGM attenuation using the formula of
\citet{Madau95}, which is based on the observed statistics of neutral
hydrogen absorbers seen in quasar spectra.

We compute the Ly$\alpha$ luminosities of galaxies by the following
procedure: (i) The model calculates the integrated stellar spectrum of
the galaxy as described above, based on its star formation history,
and including the effects of the distribution of stellar metallicities
and of variations in the IMF. (ii) We compute the rate of production
of Lyman continuum (Lyc) photons by integrating over the stellar
spectrum, and assume that all of these ionizing photons are absorbed
by neutral hydrogen within the galaxy. We assume photoionization
equilibrium applies within each galaxy, producing Ly$\alpha$ photons
according to case~B recombination \citep[e.g.][]{Osterbrock}. We note
that for solar metallicity, 11 times as many Lyc and Ly$\alpha$
photons are produced per unit mass of stars formed for our top-heavy
(burst) IMF as compared to our solar neighbourhood (disk) IMF. (iii)
The observed Ly$\alpha$ flux or luminosity of a galaxy depends on the
fraction $f_{\rm esc}$ of Ly$\alpha$ photons which escape from the
galaxy. Ly$\alpha$ photons are resonantly scattered by neutral
hydrogen, and absorbed by dust. Early estimates of this process
\citep[e.g.][]{Charlot91} showed that only a tiny fraction of
Ly$\alpha$ photons should escape from a static neutral galaxy ISM if
even a tiny amount of dust is present. Many star-forming galaxies are
nonetheless observed to have significant Ly$\alpha$ luminosities
\citep[e.g.][]{Kunth98, Pettini01}, and this is generally ascribed to
the presence of galactic winds in these systems, which allow
Ly$\alpha$ photons to escape after many fewer resonant
scatterings. Radiative transfer calculations of Ly$\alpha$ through
winds have shown that this process can explain the asymmetric
Ly$\alpha$ line profiles which are typically observed
\citep[e.g.][]{Ahn04}. The effects of radiative transfer of
Ly$\alpha$ through clumpy dust and gas have been considered by
\citet{Neufeld91} and \citet{Hansen05}.

Calculating Ly$\alpha$ escape fractions from first principles is
clearly very complicated, and so we instead adopt a simpler
approach. In Paper~I, we found that assuming a fixed escape fraction
$f_{\rm esc}$ for each galaxy, regardless of its dust properties,
resulted in a surprisingly good agreement between the predicted number
counts of emitters and the available observations. In that paper, we
chose $f_{\rm esc}=0.02$ to match the number counts at $z\approx 3$ at
a flux $f\approx 2\times 10^{-17} \ergs$, and we use the same value of
$f_{\rm esc}$ in this paper. Although this extreme simplification of a
constant escape fraction may seem implausible, it does give a
reasonably good match to the observed Ly$\alpha$ luminosity functions
and equivalent widths at different redshifts, as we show in the next
sections.

Our calculations do not include any attenuation of the Ly$\alpha$ flux
from a galaxy by propagation through the IGM. Ly$\alpha$ photons can
be scattered out of the line-of-sight by any neutral hydrogen in the
IGM close to the galaxy. If the emitting galaxy is at a redshift
before reionization, when the IGM was still mostly neutral, this could
in principle strongly suppress the observed Ly$\alpha$ flux
\citep{Miralda98}. However, various effects can greatly reduce the
amount of attenuation: ionization of the IGM around the galaxy
\citep{Madau00, Haiman02}, clearing of the IGM by galactic winds,
gravitational infall of the IGM towards the galaxy, and redshifting of
the Ly$\alpha$ emission by scattering in a wind \citep{Santos04a}. In
any case, since measurements of Gunn-Peterson absorption in quasars
show that reionization must have occured at $z\gsim 6.5$, attenuation
of Ly$\alpha$ fluxes by the IGM should not affect our predictions for
$z\lsim 6.5$, but only our predictions for very high redshifts given
in Section~\ref{sec:high-z}.

\begin{figure}
{\epsfxsize=8.5truecm
\epsfbox{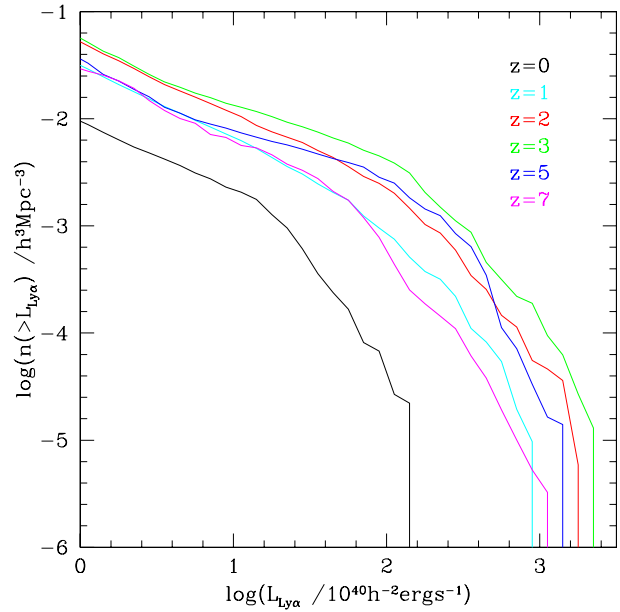}}
\caption
{ The predicted evolution with redshift of the cumulative Ly$\alpha$
luminosity function, defined as the comoving number density of
galaxies with Ly$\alpha$ luminosities brighter than
$L_{Ly\alpha}$. The model predictions are shown for selected redshifts
in the interval $z=0$ to $z=7$.  }
\label{fig:LF_lowz}
\end{figure}

\begin{figure*}
{\epsfxsize=16.truecm
\epsfbox{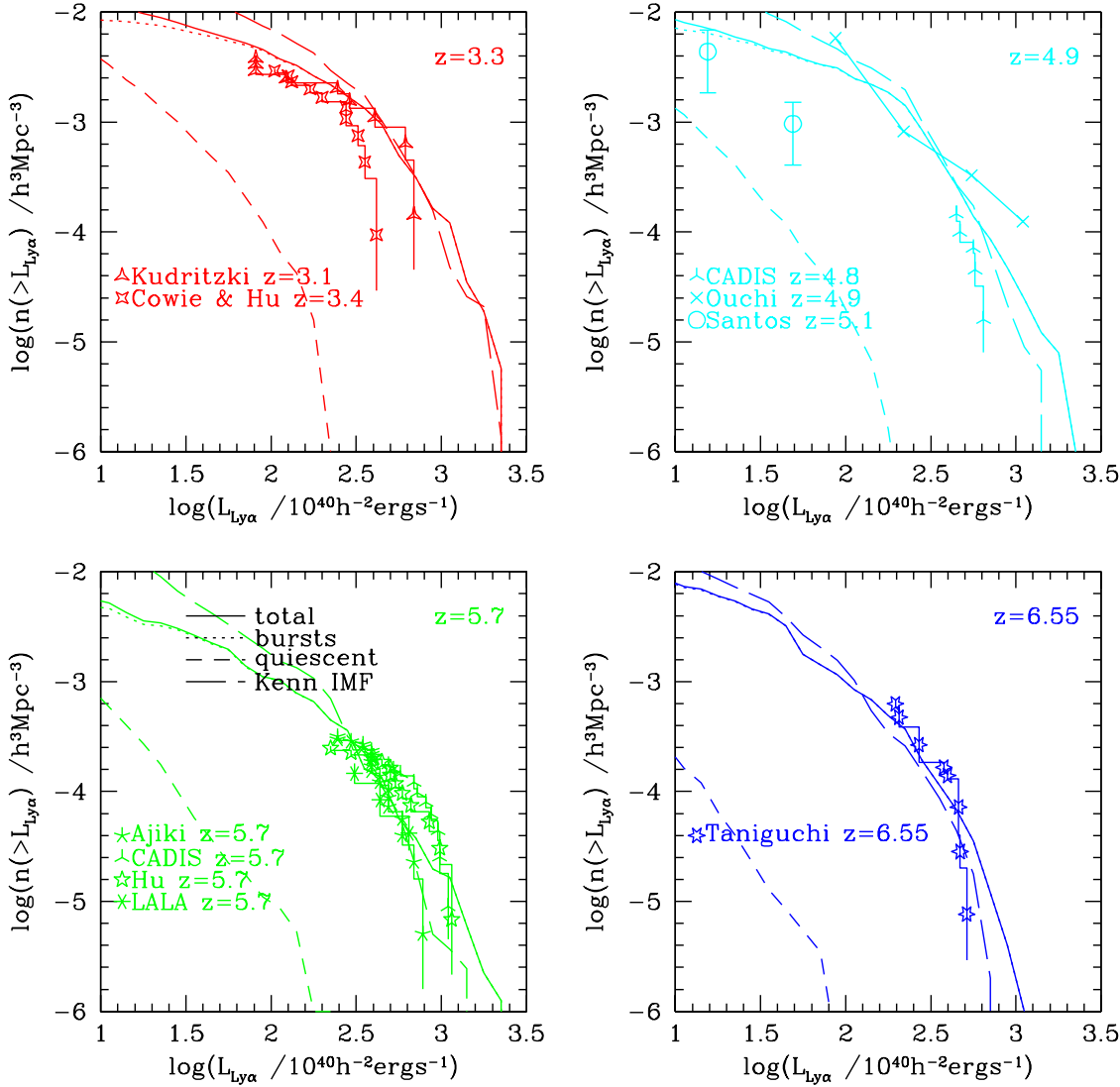}}
\caption
{ The evolution of the cumulative Ly$\alpha$ luminosity function with
redshift, comparing models with observational data. Each panel
corresponds to a different redshift, as indicated by the legend. The
curves show model predictions, while the lines with symbols and
symbols with error bars show observational data. The solid curves show
the predictions for the total luminosity function for our standard
model, with a top-heavy IMF in bursts and $f_{\rm esc}=0.02$, while
the dotted and short-dashed lines show the separate contributions of
starbursts and quiescently star-forming galaxies respectively (in most
cases the solid and dotted lines overlap). The long-dashed curves show
the predicted total luminosity function for a variant model with a
universal Kennicutt IMF and $f_{\rm esc}=0.2$.  The references for the
observational data (as shown in the symbol key) are as follows:
Kudritzki - \citet{Kudritzki00}; Cowie \& Hu - \citet{CH98}; CADIS -
\citet{Maier03}; Ouchi - \citet{Ouchi03}; Santos - \citet{Santos04b};
Ajiki - \citet{Ajiki03}; Hu - \citet{Hu04}; LALA - \citet{Rhoads03};
Taniguchi - \citet{Taniguchi05}. The redshifts for the observational
data are close to that of the model plotted in each panel, but do not
exactly coincide in all cases.  In most cases, the data are plotted
as stepped histograms, with each step corresponding to a single
galaxy. }
\label{fig:LF_4panel}
\end{figure*}

\section{Evolution of the Ly$\alpha$ luminosity function}
\label{sec:lf-evoln}

A basic prediction of our model is the evolution of the luminosity
function of Ly$\alpha$ emitters with redshift. This depends on the
distribution of star formation rates in quiescent and starburst
galaxies (with solar neighbourhood and top-heavy IMFs respectively),
and on the metallicity with which the stars are formed. Paper~I showed
predictions for the cumulative number counts of emitters per unit
redshift as a function of observed Ly$\alpha$ flux. Here we focus on a
closely related quantity, the cumulative space density of emitters as
a function of Ly$\alpha$ luminosity at different redshifts.
Fig.\ref{fig:LF_lowz} shows the cumulative luminosity function of
Ly$\alpha$ emitters predicted by our standard model for a set of
redshifts over the interval $z=0-7$. The model luminosity function
initially gets brighter with increasing redshift, peaking at $z=3$,
before declining again in number density at even higher redshifts. The
increase in the luminosity function from $z=0$ to $z\sim 3$ is driven
both by the increase in galaxy star formation rates, and by the
increasing fraction of star formation occuring in bursts (which have a
top-heavy IMF). As shown in Fig.1 in \citet{Baugh05}, the model
predicts that the fraction of all star formation occuring in bursts
increases from $\sim 5\%$ at $z=0$ to $50\%$ at $z\sim 3.5$ and then
to $\sim 80\%$ at $z\gsim 6$.

We compare the model predictions with observational estimates of the
cumulative Ly$\alpha$ luminosity function in Fig.\ref{fig:LF_4panel},
where we show different redshifts in different panels. The
observational estimates of the luminosity functions which we plot have
been calculated by \citet{Tran04} (and also S.~Lilly, private
communication, for $z=6.55$) from published data on surveys for LAEs,
assuming the same cosmology as we assume in our models. The observed
luminosity functions in the figures are labelled according to the
survey from which they were obtained. In each case, surveys using
narrow-band filters were used to find candidates for LAEs at
particular redshifts, and then either broad-band colours or follow-up
spectroscopy were used to determine which of the candidates were
likely to be real Ly$\alpha$ emitters, and which were likely to be
lower-redshift interlopers resulting from other emission lines falling
within the narrow-band filter response. The surveys with spectroscopic
follow-up which we plot are \citet{Kudritzki00}, \citet{CH98},
\citet{Hu04}, \citet[][LALA]{Rhoads03} and \citet{Taniguchi05}, while
the surveys using only colour selection are \citet[][CADIS]{Maier03},
\citet{Ouchi03} and \citet{Ajiki03}. We also show the data of
\citet{Santos04b} at $z \approx 5$ from a spectroscopic survey of
gravitationally-lensed fields.  The stepped appearance of most of the
observed luminosity functions results from the small number of objects
in most of the samples; each step corresponds to the inclusion of an
additional object as the luminosity is reduced. The cumulative
luminosity functions cut off at the bright end where the observational
samples contain only one object of that luminosity; the statistical
uncertainties are correspondingly largest at the highest
luminosities. For reference, we note that the luminosity distance for
our assumed cosmology is $d_L = (2.0,3.2,3.8,4.5)\times 10^4
h^{-1}\Mpc$ for $z=3.3, 4.9, 5.7, 6.55$ respectively, so that a
Ly$\alpha$ flux of $10^{-17}\ergcms$ at each of these redshifts
corresponds to a luminosity $(0.5,1.2,1.7,2.4)\times 10^{42}
h^{-2}\ergs$ respectively. The lower luminosity limits on the observed
luminosity functions correspond to roughly the same Ly$\alpha$ flux
limit $\sim 10^{-17}\ergcms$ at each redshift.

In Fig.\ref{fig:LF_4panel}, the predictions for our standard
model (with a top-heavy IMF in bursts and $f_{\rm esc}=0.02$) are
shown by solid lines.  We also show the separate contributions of
bursting and quiescently star-forming galaxies respectively as dotted
and short-dashed lines. In most cases, the dotted line is barely
distinguishable from the solid line, showing that the model Ly$\alpha$
luminosity function is completely dominated by bursts over the range
of redshift and luminosity plotted in
Fig.\ref{fig:LF_4panel}. Overall, there is broad agreement between the
predicted and observed luminosity functions over the redshift range
$z=3-6.6$. This is remarkable, since we allowed ourselves to adjust
only one model parameter to fit the observational data, namely the
Ly$\alpha$ escape fraction $f_{\rm esc}$. The model luminosity
functions do not perfectly match all of the observational data, but
where there are differences between the model and observational data,
there are also equally large differences between different
observational datasets. The differences between different
observational datasets could be due to a combination of (a)
statistical fluctuations (most of the samples are small), (b)
field-to-field variance due to galaxy clustering
\citep[e.g.][]{Ouchi03, Shimasaku03}, (c) differences in the details
of how the samples are selected (e.g. differences in the equivalent
width limit or photometric criteria applied), and (d) differing levels
of contamination by objects which are not Ly$\alpha$ emitters. We
note that the model predictions shown in Fig.\ref{fig:LF_4panel} do
not include any limit on the equivalent width (EW) of the Ly$\alpha$
emission line, while the observational data shown all incorporate
different lower limits on the EW of line emission as well as on the
line flux. However, as we show in \S\ref{sec:EW}, these EW thresholds
are predicted not to significantly affect the comparison of model and
observed luminosity functions in Fig.\ref{fig:LF_4panel}.

The value of the Ly$\alpha$ escape fraction which we find fits the
data, $f_{\rm esc}=0.02$, is quite small. This is mainly because, over
the range of redshift and luminosity probed by the observations, the
counts of objects in our standard model are dominated by bursts,
and we have assumed a top-heavy IMF in bursts. As noted above, the
Ly$\alpha$ luminosity for a given star formation rate is about 10
times larger with the top-heavy IMF than with a solar neighbourhood
IMF. We also show in Fig.\ref{fig:LF_4panel} by long-dashed lines
the predictions of a variant model, in which we assume the same
Kennicutt IMF for bursts and quiescent star formation, and with
$f_{\rm esc}=0.2$. Even though we have chosen $f_{\rm esc}$ for this
variant model to provide the best overall match to the observational
data in Fig.\ref{fig:LF_4panel}, we see that it agrees somewhat less
well with the data than does our standard model, especially at
$z=3.3$, where it predicts more low-luminosity galaxies. Moreover,
this variant model dramatically under-predicts both the counts of
sub-mm galaxies and the number of Lyman-break galaxies (see Fig.5(a)
in \citet{Baugh05}).


We have also investigated the effect on the predicted luminosity
functions of changing the IGM reionization redshift from our standard
value $z_{\rm reion}=10$. As described in \S\ref{sec:GALFORM}, this
affects galaxies in our model through photo-ionization feedback. For
our standard model, we find that varying $z_{\rm reion}$ over the
range 6.5 to 20 changes the luminosity function by less than the
scatter between different observational datasets in
Fig.\ref{fig:LF_4panel}, over the range of luminosity and redshift
probed by those data. Choosing a different value of $z_{\rm reion}$ in
this range would therefore not significantly affect any of the
conclusions we draw here.

\citet{Furlanetto05} have computed luminosity functions of
Ly$\alpha$ emitters from a numerical simulation, including emission
from gas heated by shocks and by the intergalactic ionizing background
as well as emission from star-forming regions in galaxies. They assume
that stars all form with a Salpeter IMF. However, the luminosity
functions which they compute combine all of the emission from each
dark matter halo, and so are different from the luminosity functions
of individual galaxies which our model predicts. Furthermore, they
effectively assume an escape fraction $f_{\rm esc}=1$ for Ly$\alpha$
emission from star formation. They do not make any detailed comparison
with observational data on Ly$\alpha$-emitting galaxies, but note that
their luminosity functions predict roughly an order-of-magnitude more
objects than are observed over the range $L_{Ly\alpha} \sim 10^{42} -
10^{43} \ergs$. This is roughly consistent with what we would find if
we assumed a Kennicutt IMF for all star formation and $f_{\rm
esc}=1$.

\section{Observable properties of Ly$\alpha$ emitters }
\label{sec:obs-props}

Now that we have established that our model gives a very good match 
to the luminosity function of Ly$\alpha$ emitters at different redshifts, 
we turn our attention to other observable properties of these objects. 
We first present predictions for the distribution of Ly$\alpha$ equivalent 
widths (\S4.1), before examining the broad-band continuum magnitudes of 
Ly$\alpha$ emitters (\S4.2) and, finally, the size distribution 
of emitters (\S4.3).

\subsection{Ly$\alpha$ equivalent widths}
\label{sec:EW}

Our model allows a simple prediction for the equivalent width (EW) of
the Ly$\alpha$ emission line in each galaxy: we divide the luminosity
in the emission line by the mean luminosity per unit wavelength of the
stellar continuum on either side of the line. We distinguish between
the {\em net} and {\em intrinsic} line and stellar luminosities and
equivalent widths. The {\em net} values are obtained after we multiply
the Ly$\alpha$ luminosity by the escape fraction $f_{\rm esc}$ and
after we attenuate the stellar luminosity by dust extinction, while
the {\em intrinsic} values are those before we include either the
Ly$\alpha$ escape fraction or dust attenuation. A limitation of our
current model is that it does not include the effects of absorption of
Ly$\alpha$,
so the equivalent widths we calculate are always positive (or
zero). Ly$\alpha$ absorption features
(corresponding to negative equivalent widths) could be produced either
by absorption in stellar atmospheres \citep{Charlot93}, or by neutral
gas within the galaxy or in an expanding shell or wind around it
\citep{Tenorio99}. Our calculations are therefore incomplete, but
nevertheless represent an important first step.

\begin{figure}
\centering

{\epsfxsize=6truecm
\epsfbox{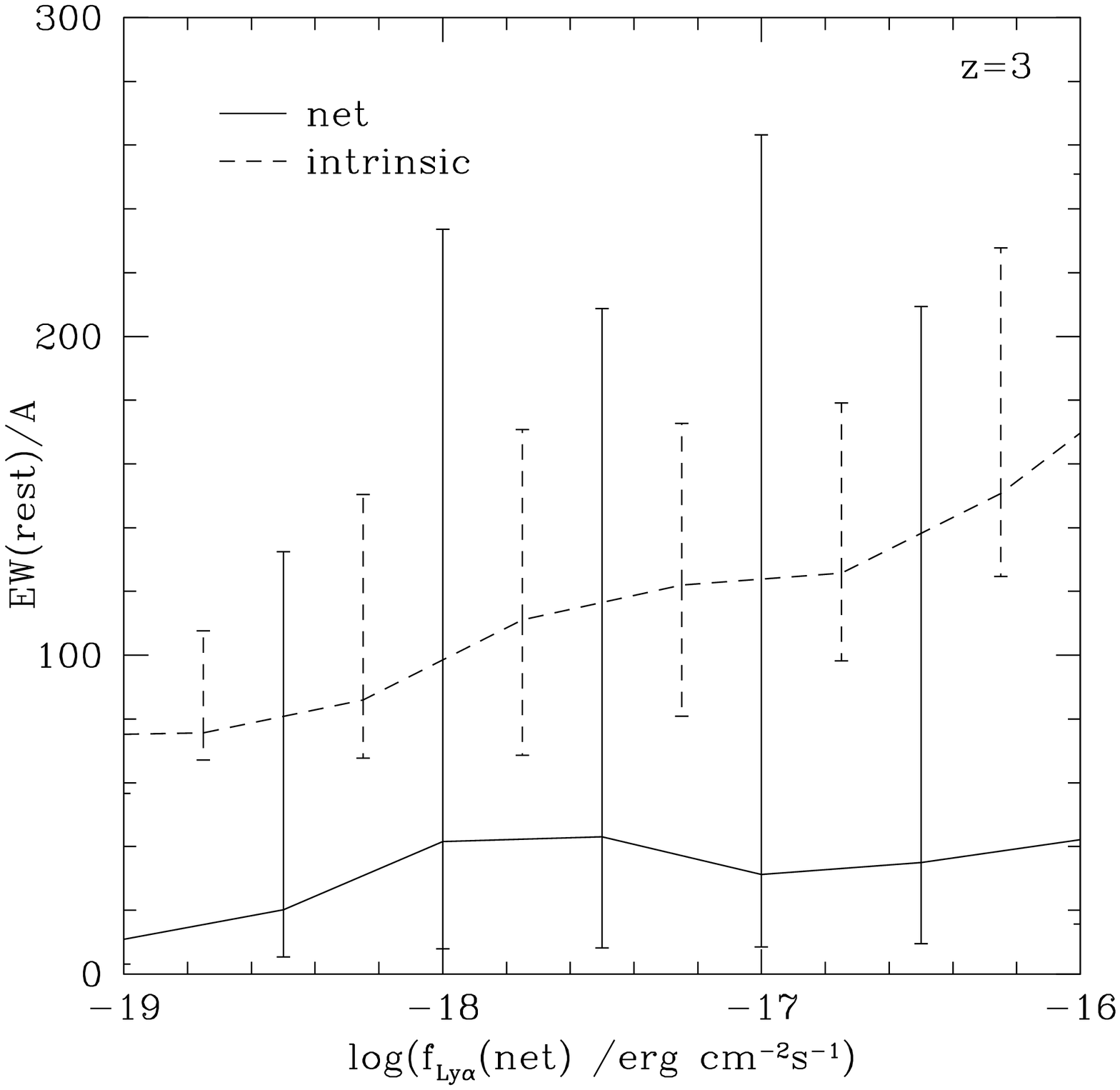}}
{\epsfxsize=6truecm
\epsfbox{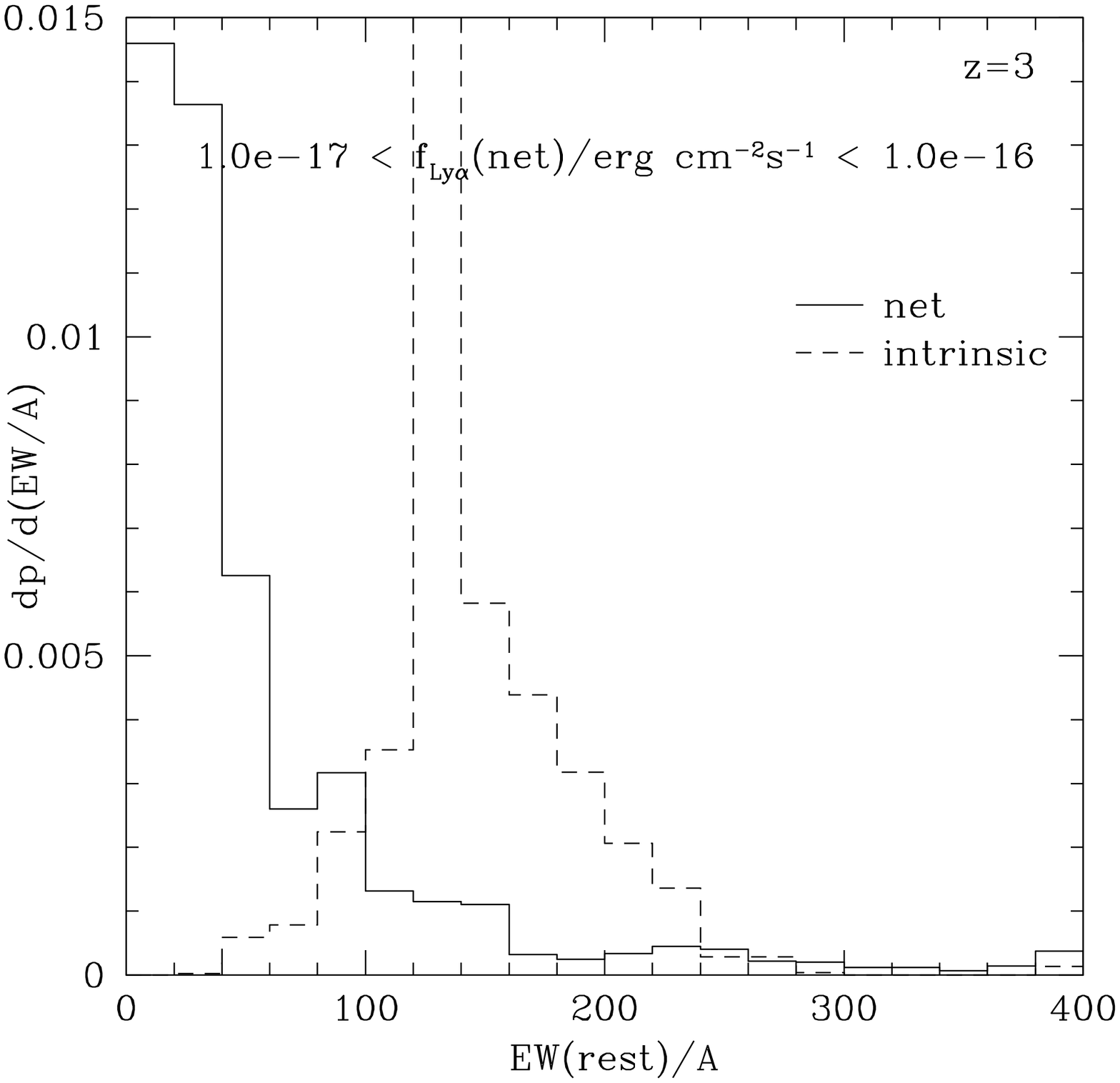}}
{\epsfxsize=6truecm
\epsfbox{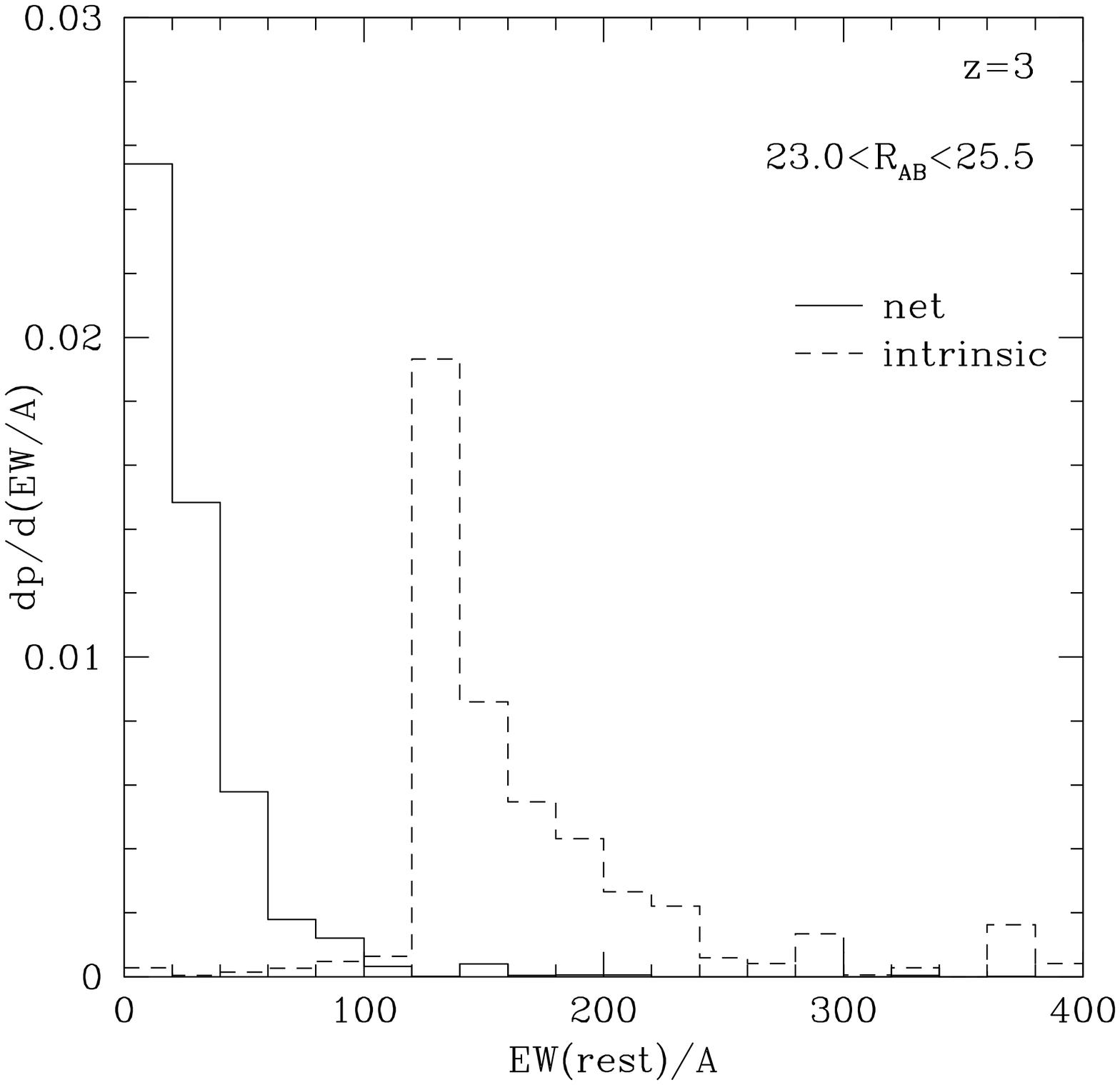}}

\caption{ The predicted rest-frame equivalent width (EW) of the
  Ly$\alpha$ emission line for galaxies at $z=3$. We show results both
  for the intrinsic EW, i.e. before including attenuation by neutral
  gas and dust in the galaxy, and for the net EW, i.e. after including
  the escape fraction for Ly$\alpha$ photons and dust extinction of
  the stellar continuum - these are shown by dashed and solid lines
  respectively. In either case, the EW is considered as a function of
  the net Ly$\alpha$ flux. 
  (a) The upper panel shows the
  predicted median EW as a function of the net Ly$\alpha$ flux. The
  error bars show the 10-90 percentile range at a given flux. (b) The
  middle panel shows the predicted distribution of EWs for galaxies
  with net Ly$\alpha$ fluxes in the range
  $10^{-17}-10^{-16}\ergcms$. 
(c) The lower panel
  shows the predicted EW distribution for galaxies selected to have
  continuum magnitudes in the range $23<R_{AB}<25.5$.  }
\label{fig:EW}
\end{figure}

Fig.\ref{fig:EW} shows the model predictions for rest-frame equivalent
widths of Ly$\alpha$-emitting galaxies at z=3.  The most remarkable
feature of these plots is the wide spread of EWs predicted by the
model. This is seen most clearly in the middle panel, which shows the
distribution of equivalent widths for galaxies selected to have
Ly$\alpha$ fluxes in the range $10^{-17} < f < 10^{-16}\ergcms$. We
see that there is a big difference between the distributions of
intrinsic and net EWs (shown by dashed and solid lines
respectively). For this flux range, the intrinsic rest-frame EW has a
median value of 130\AA, and most galaxies have EWs in the range
100--200\AA. These values are similar to the predictions of
\citet{Charlot93}. The spread in intrinsic EWs results mostly from the
spread in burst ages and timescales. For the same galaxies, the net
EWs have a much lower median value, 33\AA, but with a much broader
distribution, with a peak close to 0 and a tail extending up to $\sim
400$\AA. Since our model assumes that all galaxies have the same
escape fraction for Ly$\alpha$, this broad distribution of net EWs
results from the wide spread in values of dust extinction for the
stellar continuum. The Ly$\alpha$ escape fraction reduces the net EW
relative to the intrinsic value, but dust extinction of the stellar
continuum increases it. 

The upper panel of Fig.\ref{fig:EW} shows the median and 10-90
percentile range for the EW of Ly$\alpha$ as a function of the net
Ly$\alpha$ flux.  There is a weak trend of EW increasing with
Ly$\alpha$ flux (or luminosity). In the case of the intrinsic EW, this
increase is driven mostly by the shift from being dominated by
quiescently star-forming galaxies (with a normal IMF) at low
luminosities to being dominated by bursts (with a top-heavy IMF) at
high luminosities, and by the change in the typical star formation
history. For the net EW, the increase in the median is driven also by
the increase in the typical dust extinction of the stellar continuum
with increasing luminosity.

The Ly$\alpha$ equivalent widths predicted by our model are similar to
those found in observed galaxy samples selected by their Ly$\alpha$
emission. \citet{CH98} and \citet{Kudritzki00} selected LAEs having
Ly$\alpha$ fluxes $\sim 10^{-17}-10^{-16} \ergcms$ at $z=3.4$ and
$z=3.1$ respectively. In both cases, their narrow-band selection
imposed a lower limit on the rest-frame EW $\approx20$\AA\ for the
detected objects, and the median rest-frame EW of the objects above
this threshold was found to be $\approx 40$\AA. This appears broadly
compatable with the predictions shown in Fig.\ref{fig:EW}, once one
allows for the fact that the EW threshold in the observed samples will
raise the median EW above the value expected in the absence of any EW
threshold. At a higher redshift, $z\approx 4.5$, \citet{Dawson04}
selected LAEs with $f\sim10^{-17}-10^{-16} \ergcms$ and $EW({\rm
rest})> 15$\AA, and measured a median $EW({\rm rest}) \approx 80$\AA\
for their sample. This is also in good agreement with our model, which
predicts a median $EW({\rm rest}) \approx 80$\AA\ for LAEs with
$10^{-17} < f < 10^{-16}\ergcms$ at this redshift.

\begin{figure}
\centering

{\epsfxsize=8truecm
\epsfbox{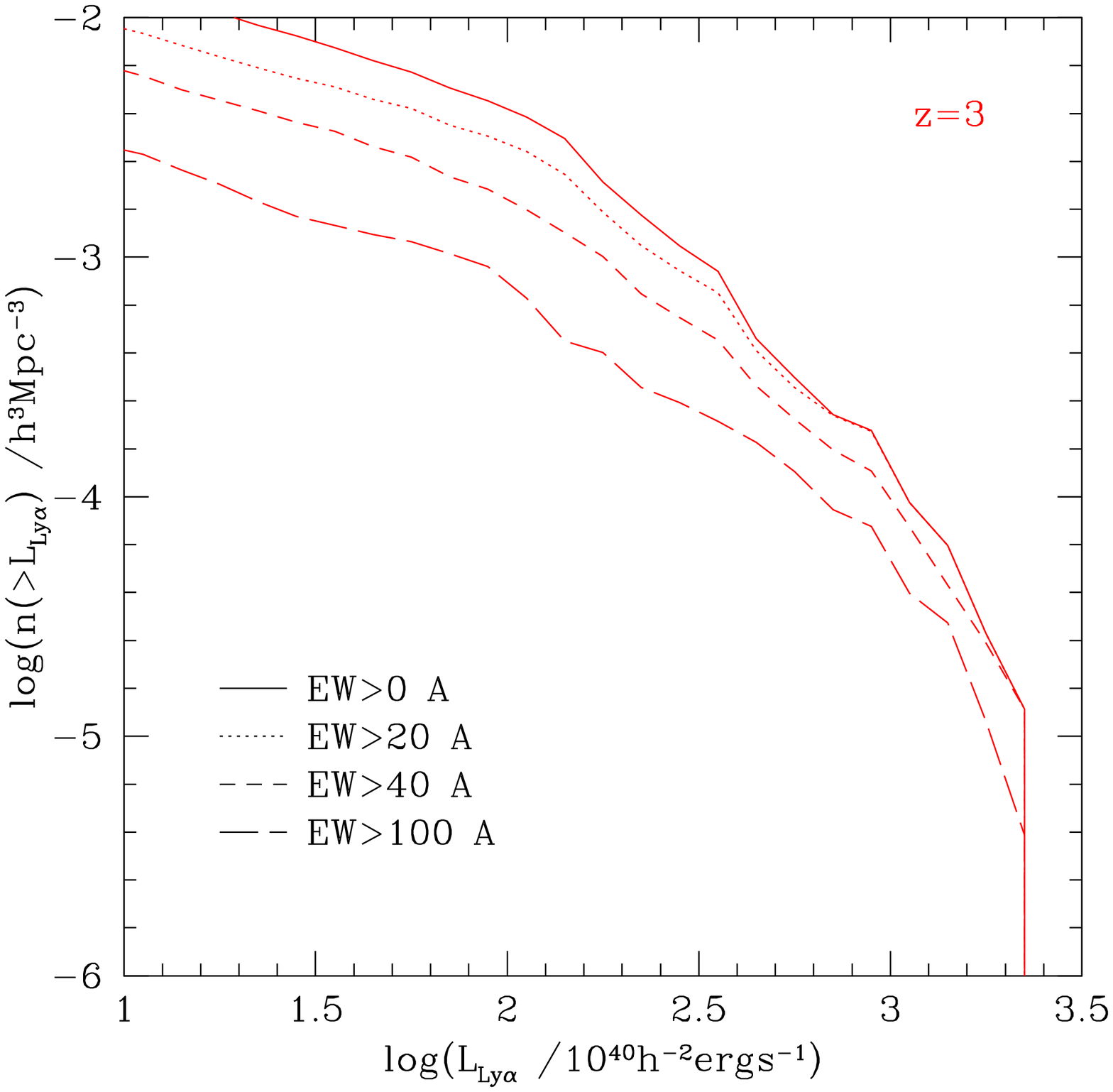}}
{\epsfxsize=8truecm
\epsfbox{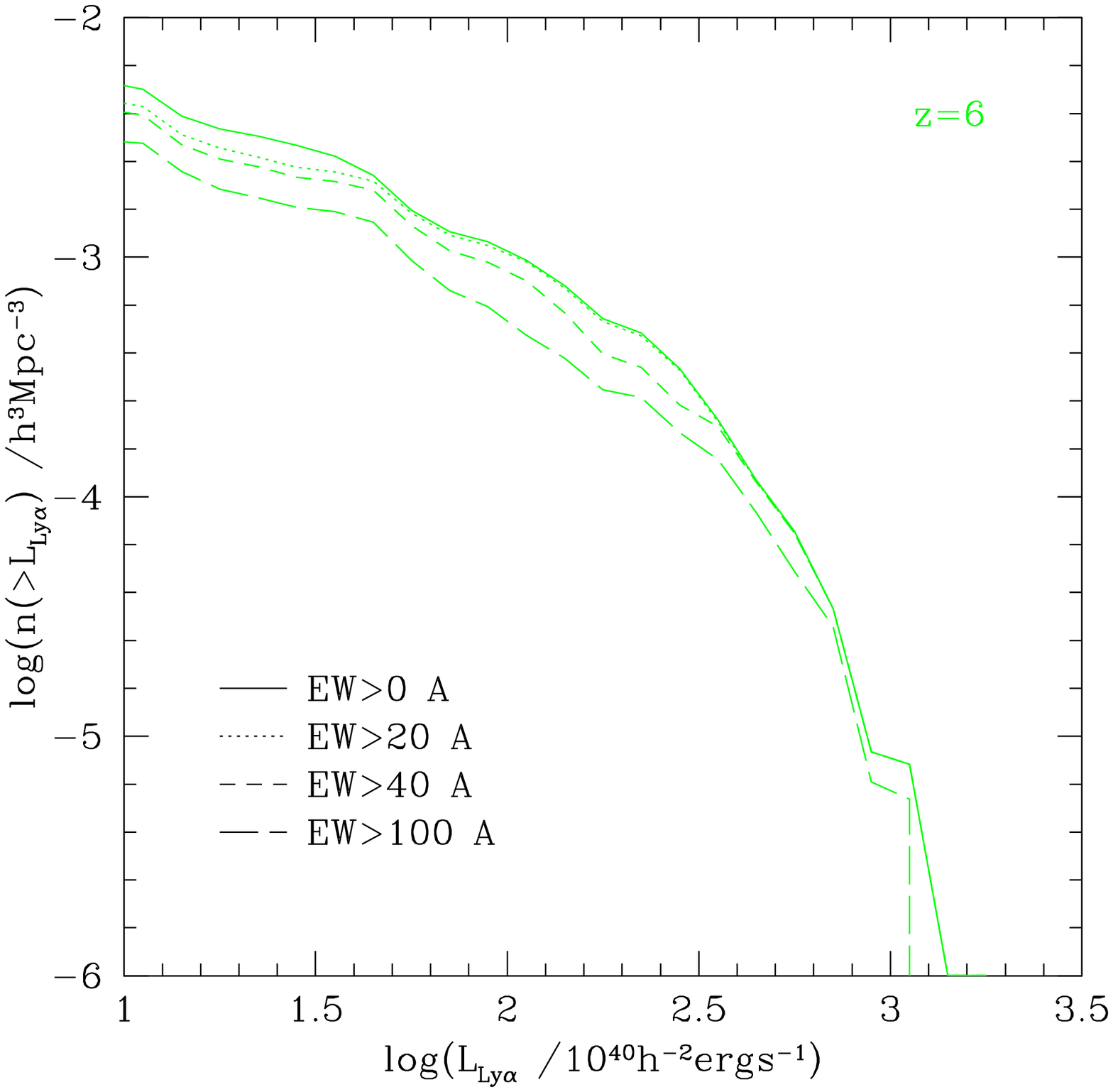}}

\caption{The effect of different equivalent width thresholds on the
  predicted luminosity function of Ly$\alpha$ emitters. In each panel,
  the lines show the predicted cumulative luminosity function for
  different lower limits on the rest-frame EW of Ly$\alpha$ emission:
  EW$>0$\AA\ (solid curves), EW$>20$\AA\ (dotted), EW$>40$\AA\
  (short-dashed), and EW$>100$\AA\ (long-dashed). (a) $z=3$. (b)
  $z=6$. }
\label{fig:LF_EW}
\end{figure}

\citet{Shapley03} have measured Ly$\alpha$ emission and absorption
profiles and EWs in a sample of galaxies at $z\sim 3$ selected using
the Lyman-break technique. Their sample is thus selected on rest-frame
far-UV stellar continuum luminosity, rather than on the presence of a
strong Ly$\alpha$ emission line. They find that $\sim 30\%$ of their
galaxies show Ly$\alpha$ only in emission, $\sim 30\%$ of galaxies
show Ly$\alpha$ only in absorption, and $\sim 40\%$ show a combination
of Ly$\alpha$ absorption and emission.  They find a very asymmetric
and skewed distribution of Ly$\alpha$ rest-frame EW's, with a median
close to 0\AA, extending down to $\sim -50$\AA\ for net absorption and
to $\gsim 100$\AA\ for net emission. For the galaxies with net
Ly$\alpha$ emission, the median EW is $\sim 20$\AA. (For galaxies with
net absorption, the median EW is $\sim -20$\AA.) The lower panel of
Fig.\ref{fig:EW} shows the EW distribution predicted by the model if
we select galaxies in a similar way to \citeauthor{Shapley03}, with a
continuum magnitude range $23<R_{AB}<25.5$ (including dust extinction)
and no condition on the Ly$\alpha$ flux or EW. The model predicts a
median EW $\approx 20$\AA\ for this case, very similar to the typical
EW of the emission component of Ly$\alpha$ in the
\citeauthor{Shapley03} sample. The shape of the model EW distribution
(which is restricted to EW $\geq 0$) is also very similar to that
found by \citeauthor{Shapley03} for $EW>0$ (see their Fig.8). However,
without including a calculation of Ly$\alpha$ absorption in our model,
we cannot make a more detailed comparison with
\citeauthor{Shapley03}. Since a calculation of Ly$\alpha$ absorption
requires a treatment of radiative transfer through the galaxy ISM, we
defer this to a future paper.

Since our model allows us to estimate Ly$\alpha$ EWs, we can also
estimate the effect on the Ly$\alpha$ luminosity function of imposing
different lower limits on the EWs of Ly$\alpha$ emission from
galaxies. This is shown in Fig.\ref{fig:LF_EW}, for rest-frame EW
thresholds $EW_{\rm min}({\rm rest}) = 0, 20, 40$ and $100$\AA, for
redshifts $z=3$ and $z=6$. Different observational surveys for LAEs
impose different lower limits on the EWs of the objects they
include. However, for most of the observational data plotted in
Fig.\ref{fig:LF_4panel}, the lower limit is around $EW_{\rm min}({\rm
rest}) = 20$\AA. We see from Fig.\ref{fig:LF_EW} that an EW threshold
around this value is predicted to have only a small effect on the
Ly$\alpha$ luminosity function, so the conclusions we drew from the
comparison with observational data in Fig.\ref{fig:LF_4panel} would
not be significantly affected.

\begin{figure}
\centering

{\epsfxsize=7.5truecm
\epsfbox[40 170 260 700]{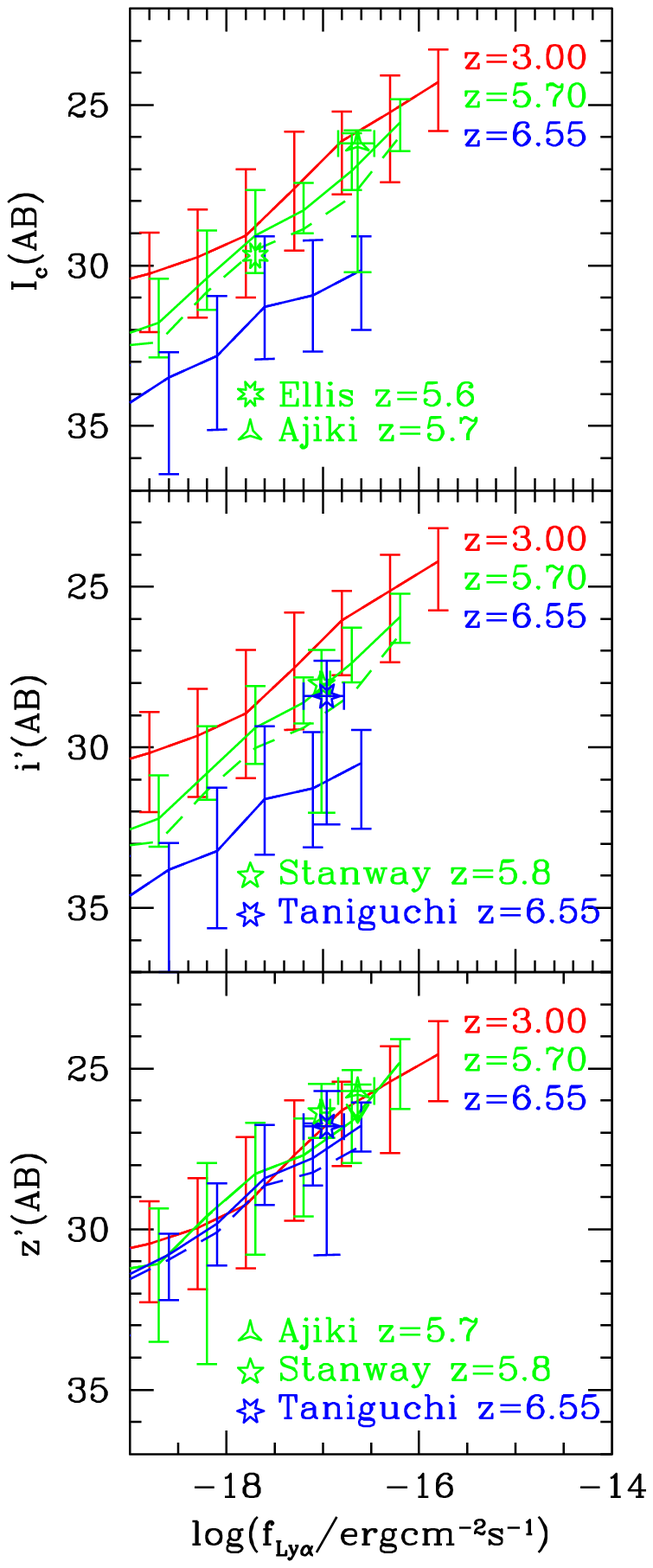}}

\caption{ Broad-band magnitudes as a function of Ly$\alpha$ flux. The
lines show the model predictions for three different redshifts, $z=3$,
5.7 and 6.55, in different colours. The dashed lines show the median
magnitude due to the stellar continuum only, while the solid lines
include also the contribution of the Ly$\alpha$ line to the broad-band
magnitude. Where the dashed line is not visible, it coincides
with the solid line.  The error bars on the lines show the 10-90
percentile range. For clarity, small offsets in the $x$-direction
have been applied to the model relations for different redshifts.
The top, middle and lower panels show results for the $I_c$,
$i^{\prime}$ and $z^{\prime}$ filters respectively. The symbols show
observational data, plotted in the same colours as the model curve
closest in redshift. The observational data are as follows: Ellis -
\citet{Ellis01} (1 galaxy); Ajiki - \citet{Ajiki03} (20 galaxies);
Stanway - \citet{Stanway04} (3 galaxies); Taniguchi -
\citet{Taniguchi05} (9 galaxies). For samples with $>1$ galaxy, we
plot an estimate of the median Ly$\alpha$ flux and broad-band
magnitude, and of the 10-90 percentile ranges in both (shown by error
bars).  }
\label{fig:mags}
\end{figure}

\subsection{Broad-band magnitudes}
Another important test for our model of the Ly$\alpha$ emitters is
that it should predict the correct stellar continuum as measured in
broad-band filters. Fig.\ref{fig:mags} shows the model predictions for
the median broad-band magnitudes as a function of Ly$\alpha$ flux at
three different redshifts, $z=3$, 5.7 and 6.55, and for three
different broad-band filters, the $I_c$, $i^{\prime}$ and $z^{\prime}$
filters on the Suprime Cam on the Subaru Telescope. (We chose these
particular filters because most of the observational data we will
compare with were taken with them.) The predicted broad-band
magnitudes include the effects of dust extinction and of attenuation
by the intervening IGM (based on \citealt{Madau95}). The
evolution with redshift of the predicted $I_c$ and $i^{\prime}$
magnitudes at a given Ly$\alpha$ flux which is seen in
Fig.\ref{fig:mags} results mostly from the IGM opacity.  In some
cases, the Ly$\alpha$ emission line falls within the bandpass of the
filter. We have therefore computed broad-band magnitudes due to either
the stellar continuum only (shown by dashed lines), or to the stellar
continuum and Ly$\alpha$ emission line together (shown by solid
lines). In most of the cases plotted in Fig.\ref{fig:mags}, the solid
and dashed lines are indistinguishable, but in a few cases there is a
small offset, showing that the Ly$\alpha$ line makes a modest
contribution to the broad-band magnitude in these cases.

For comparison, we also plot in Fig.\ref{fig:mags} a selection of
observational data for galaxies at $z\sim 5-7$ from the following
papers: \citet{Ellis01}, \citet{Ajiki03}, \citet{Stanway04} and
\citet{Taniguchi05}. The data we plot constrain the stellar continuum
at wavelengths $\sim 900-1400$\AA\ in the galaxy rest-frame. The
\citeauthor{Ellis01} and \citeauthor{Stanway04} data were actually
taken on HST using the WFPC2 ($I_{814}$ filter) and ACS ($i'$ and $z'$
filters) cameras respectively, but we have verified that the
difference of these filters from the Subaru filters (in particular,
the difference of $I_{814}$ from $I_c$) does not significantly affect
the comparison of models with data which we make here. In cases where
the observational papers have tried to correct the broad-band
magnitudes for the contribution from the Ly$\alpha$ line, we have
plotted the total magnitude before this correction was made. We plot
the observational data in Fig.\ref{fig:mags} as symbols of the same
colour as the model curve closest in redshift. Apart from
\citeauthor{Ellis01}, all of the samples contain more than one galaxy,
and in these cases we estimate the median and 10-90\% percentile range
for both the Ly$\alpha$ flux and the broad-band magnitude (allowing
for upper limits on the broad-band fluxes). We plot the symbol at the
median value and show the 10-90\% range by error bars. If the 10\%
value is an upper limit, we indicate this by a very long downwards
error bar.

We see from Fig.\ref{fig:mags} that there is mostly good agreement
between the predicted broad-band magnitudes and the observational
data. The one exception is that the median $i'$ magnitude measured
from \citet{Taniguchi05} for galaxies at $z=6.55$ is nearly 3
magnitudes brighter than what our model predicts, even though the $z'$
magnitudes for the same observational sample agree very well with the
model predictions. However, 7 out of 9 objects in the
\citeauthor{Taniguchi05} sample are detected in the $i'$ band at less
than $2\sigma$ significance, so it is possible that the symbol marking
our estimate of their median magnitude is biased high by statistical
errors.  We note that at redshift $z=6.55$, the $i'$-band flux is
sensitive to emission at wavelengths $\sim 900-1100$\AA\ in the galaxy
rest-frame, so it is expected to be greatly attenuated by Ly$\alpha$
absorption by neutral hydrogen in the intervening IGM. In contrast,
the flux in the $z'$ band at this redshift is expected to be much less
affected by IGM attenuation. Thus, an alternative possible explanation
for the disagreement between the model and the
\citeauthor{Taniguchi05} $i'$ data is that we over-estimate the degree
of IGM attenuation at this redshift when we calculate it using the
\citet{Madau95} formula.

\begin{figure}
\centering

{\epsfxsize=8.5truecm
\epsfbox{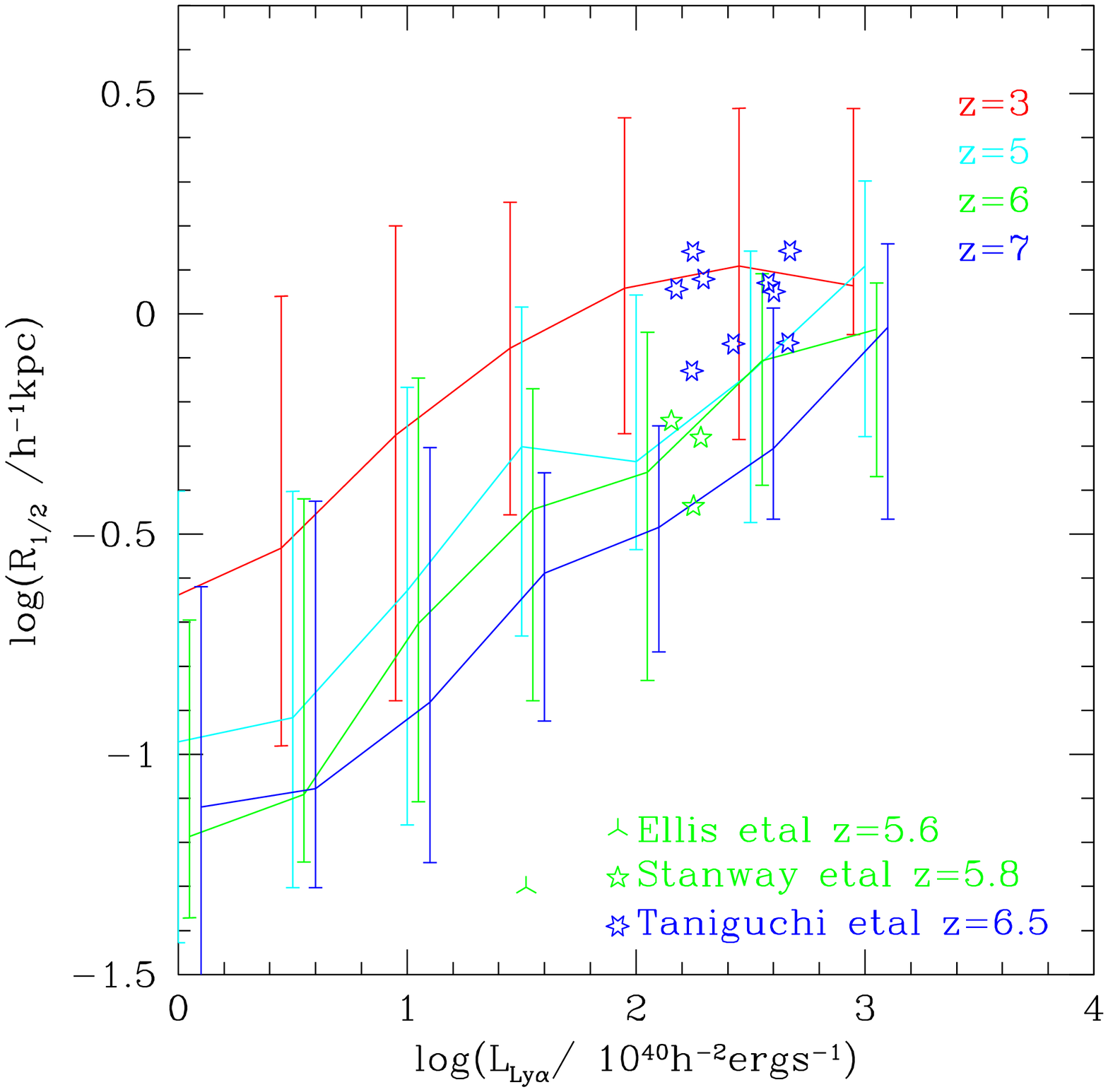}}
{\epsfxsize=8.5truecm
\epsfbox{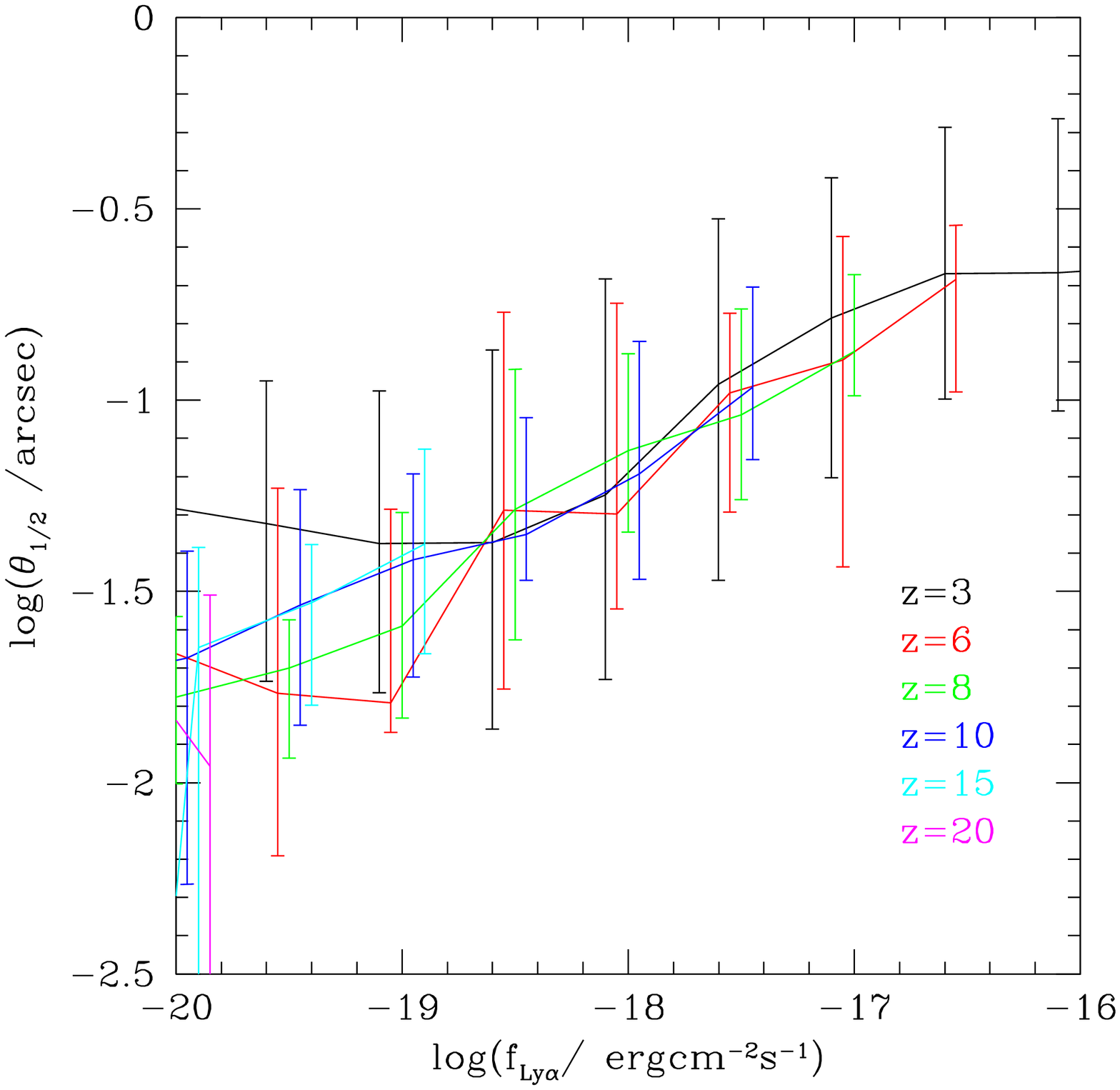}}

\caption{ The sizes of Ly$\alpha$ emitters. (a) The top panel shows
the physical sizes of Ly$\alpha$ emitters as a function of their
Ly$\alpha$ luminosities. The lines show model predictions for the
median stellar half-mass radius for four different redshifts, $z=3$,
5, 6 and 7. The error bars show the 10-90 percentile range at a given
luminosity. The symbols show observational estimates of galaxy radii,
from \citet{Ellis01}, \citet{Stanway04} and \citet{Taniguchi05},
plotted for individual galaxies in the same colours as the model curve
closest in redshift (see text for more details). (b) The lower panel
shows predicted angular sizes as a function of Ly$\alpha$ flux, for
the redshifts $z=3,6,8,10,15$ and $20$. The sizes plotted are again
stellar half-mass radii.  }
\label{fig:sizes}
\end{figure}


\subsection{Sizes of Ly$\alpha$ emitters}
Our semi-analytical model predicts the half-mass radii for the disk
and bulge components of each galaxy. From these we can compute the
half-mass radius of the stars, and also half-light radii in different
bands, allowing for different colours of the disk and bulge, but
assuming that both components have internally uniform colours. 
Coenda \etal (2005, in preparation) will present predictions from our
model for the sizes of galaxies selected by their stellar continuum
emission, and compare with observational data over the redshift range
$0<z<6$.  \citet{Cole00} have discussed predictions for galaxy sizes
at $z=0$ based on an earlier version of our semi-analytical model. In
the present paper, we will only consider the sizes of galaxies
selected to be Ly$\alpha$ emitters. We emphasize that we are not
considering in this paper the properties of Ly$\alpha$ blobs
\citep[e.g.][]{Matsuda04}, which are much more spatially extended than
typical Ly$\alpha$ emitters, and appear to be a distinct class of
object.

In Fig.\ref{fig:sizes}(a), we show model predictions for the median
stellar half-mass radius (together with its 10-90 percentile range) as
a function of Ly$\alpha$ luminosity for several different redshifts,
$z=3$, 5, 6 and 7. We have also calculated model half-light radii in
the rest-frame UV, and the results are almost identical to those for
the half-mass radius for these redshifts and luminosities. The stellar
sizes are predicted to be quite compact at these redshifts, $\lsim 1
h^{-1}\kpc$. We see that, as well as a correlation of size with
luminosity (roughly as $R_{1/2} \propto L^{1/3}$), the models also
predict that the median radius at a given luminosity should decrease
with increasing redshift (roughly as $(1+z)^{-1}$ or
$(1+z)^{-1.5}$). At a fixed Ly$\alpha$ luminosity of $10^{42.5}
h^{-2}\ergs$ (the typical value in the higher-redshift surveys shown
in Fig.\ref{fig:LF_4panel}), the median half mass radius shrinks from
$\sim 1 h^{-1}\kpc$ at $z=3$ to $\sim 0.5 h^{-1}\kpc$ at $z=7$.


We have also plotted in Fig.\ref{fig:sizes}(a) some observational
estimates of sizes for individual Ly$\alpha$-emitting galaxies at
$z\sim 5-7$, plotted in the same colour as the model curve closest in
redshift, for three different samples: \citet{Stanway04} used HST to
measure half-light radii in the rest-frame UV of 3 LAEs;
\citet{Ellis01} used HST to measure the rest-frame UV size of a single
strongly gravitationally lensed LAE; and \citet{Taniguchi05} used
ground-based narrow-band imaging to estimate the sizes of the
Ly$\alpha$-emitting region in 9 LAEs. We see that the
\citet{Stanway04} data agree very well with our model predictions, but
the \citet{Ellis01} galaxy is much smaller than the median predicted
by our model at that luminosity and redshift. However, the
\citet{Ellis01} datapoint is more uncertain than those of
\citet{Stanway04}, because it relies on the analysis of a highly
gravitationally amplified and distorted image. We also see that the
sizes of the Ly$\alpha$-emitting regions estimated by
\citet{Taniguchi05} are typically $\sim 2$ times larger than the model
prediction for the stellar half-mass radius at the same
luminosity. This might be because the Ly$\alpha$ emission in these
high-redshift LAEs really is more extended than the stellar
distribution. This has been found to be the case in some local
starburst galaxies by \citet{Mas03}, who explain this as resulting
from scattering of Ly$\alpha$ by neutral gas around these galaxies.
Alternatively, it is possible that \citeauthor{Taniguchi05} have
over-estimated the sizes of their galaxies, which are barely spatially
resolved in their ground-based images.

In Fig.\ref{fig:sizes}(b) we show predictions for the angular sizes of
Ly$\alpha$-emitters as a function of Ly$\alpha$ flux, for redshifts
over the range $z=3-20$. (We again use the stellar half-mass radius as
our measure of the size.) We see that the relation between angular
size and flux evolves rather little with redshift, even though the
relation between physical size and luminosity does evolve
appreciably. Predicted angular sizes are typically $\sim 0.1$
arcseconds for fluxes in the observed range.

\section{Predicted physical properties of Ly$\alpha$ emitters }
\label{sec:phys-props}
One of the main strengths of semi-analytical modelling lies in the
ability of the models to make predictions for a wide range of galaxy
properties.  Some of these predictions can be tested directly against
observations, as we saw in the previous section. Others can be tested
indirectly, for example through the interpretation of measurements of
clustering.  Finally, some predictions serve to illustrate how a
subset of galaxies highlighted by a particular observational selection
fit into the overall galaxy population. In this section we present
some additional predictions of the model that characterize the
Ly$\alpha$ emitters.

\begin{figure*}
{\epsfxsize=16.truecm
\epsfbox{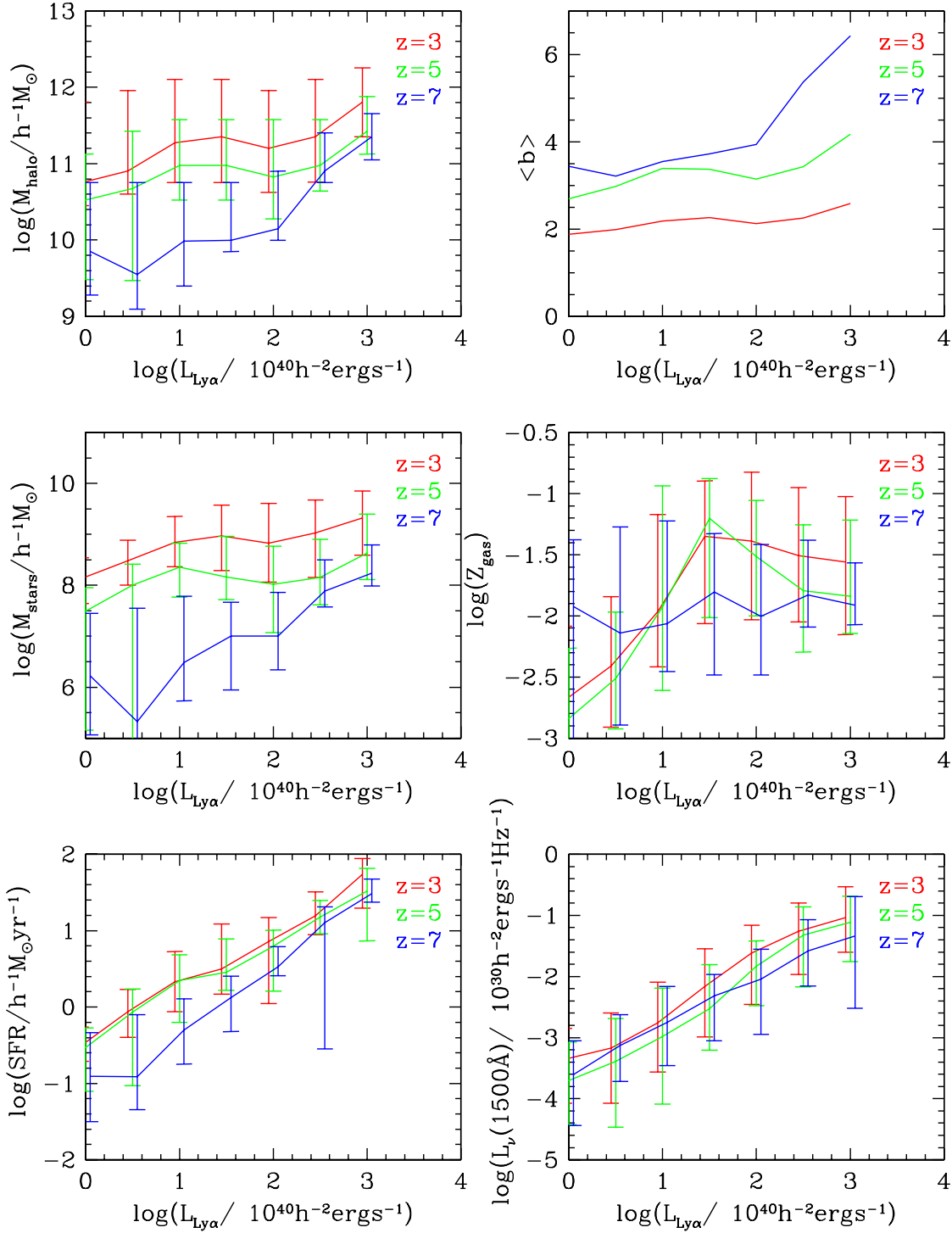}}
\caption{ Model predictions for a range of physical properties of
Ly$\alpha$ emitters plotted as a function of Ly$\alpha$ luminosity. In
each panel the model predictions are shown for $z=3,5$ and 7. The
lines show the median values of the respective properties (apart from
panel (b) which shows the mean) and the error bars show the 10-90
percentile range. (a) The upper left panel shows the mass of the
dark matter halo hosting the emitter. (b) The upper right panel shows
the mean clustering bias. (c) The middlle left panel shows the stellar
mass. (d) The middle right panel shows the metallicity of the cold
gas. (e) The lower left panel shows the total star formation rate. (f)
The lower right panel shows the UV continuum luminosity at a
rest-frame wavelength of 1500\AA.  }
\label{fig:props}
\end{figure*}

Fig.\ref{fig:props} shows model predictions for different physical
properties as a function of Ly$\alpha$ luminosity. In each panel, we
show the predictions at $z=3,5$ and 7. We plot the median value of the
respective quantity and indicate the spread in the predicted values by
showing the 10-90 percentile range, apart from the plot of clustering
bias, where we show only the mean value.

\clearpage

\subsection{Halo masses}
\label{sec:halo-mass}
The upper left panel of Fig.\ref{fig:props} shows the masses of the
dark matter halos hosting Ly$\alpha$ emitters. We see that at $z=3$,
there is only a weak dependence of median halo mass on Ly$\alpha$
luminosity, while at $z=7$, the dependence is much stronger. At a
given luminosity, the typical halo mass decreases with increasing
redshift, with this trend being stronger at lower luminosities. As
discussed in \S\ref{sec:SFR} below, the Ly$\alpha$ luminosity traces
the instantaneous star formation rate (SFR) quite well in our model,
but with a ratio which is $\sim 10$ times larger for bursting compared
to quiescent galaxies, because of the difference in IMFs.  There are
two main reasons for the weak dependence of halo mass on Ly$\alpha$
luminosity at the lower redshifts: (a) the bursts introduce a large
scatter into the relation between instantaneous SFR and object mass,
especially at lower redshifts;  (b) the shift from being dominated by
bursts at high Ly$\alpha$ luminosities to being dominated by quiescent
disks at lower luminosities flattens the trend of SFR with Ly$\alpha$
luminosity, which tends to hide the underlying trend of halo mass with
SFR.
Current Ly$\alpha$ surveys probe objects with luminosities $\sim
10^{42.5} h^{-2}\ergs$ over the whole redshift range $z\sim 3-7$, for
which the typical halo mass $\sim 10^{11} h^{-1} \Msol$, declining by
a factor $\sim 3$ from $z=3$ to $z=7$.

\subsection{Clustering bias}
\label{sec:bias}
The halo masses of LAEs can be constrained observationally from
measurements of their clustering. Since the predicted halo masses for
typical observed LAEs are larger than the characteristic halo mass at
each redshift, we expect the LAEs to be more strongly clustered than
the dark matter. We have used the halo masses predicted by the model
to calculate the linear clustering bias $b$, which is expected to
describe the clustering on large scales, using the formula of
\citet{Sheth01}. We calculate a mean bias for objects in each range of
luminosity. This mean bias is shown in the upper right panel of
Fig.\ref{fig:props}.  Over the luminosity range $L_{Ly\alpha} =
10^{40} - 10^{43} h^{-2}\ergs$, our model predicts that the bias
increases with redshift at fixed luminosity. At $z=3$, we predict
$b\approx 2$ over this luminosity range, increasing only very slightly
with luminosity. At $z=7$, the bias is predicted to vary much more
strongly with luminosity, from $b\approx 3$ at $L_{Ly\alpha} =
10^{40}h^{-2}\ergs$ to $b\approx 6$ at $10^{43} h^{-2}\ergs$. These
predictions for the clustering of LAEs seem generally consistent with
current observational constraints. The most reliable measurement of
the clustering of LAEs to date is probably that of \citet{Ouchi05},
since their sample covers by far the largest comoving volume. For
galaxies at $z=5.7$ with Ly$\alpha$ luminosities $\sim 10^{42}
h^{-2}\ergs$, they find a large-scale clustering bias $b= 3.4 \pm
1.8$. For the same redshift and luminosity, our model predicts
$b\approx 4$, in excellent agreement with this measurement. At
somewhat lower redshifts, $z\approx 4.8$, somewhat conflicting results
have been obtained for the clustering \citep[e.g.][]{Ouchi03,
Shimasaku03, Shimasaku04, Hamana04}, however, these have been obtained
from much smaller survey volumes. In particular, \citet{Shimasaku04}
measure very different clustering in their two approximately equal
survey volumes at $z=4.8$, showing that the surveys used were not
large enough to reliably measure the average clustering.

\subsection{Stellar masses}
\label{sec:stellar-mass}
The middle left panel of Fig.\ref{fig:props} shows the predicted
stellar masses of LAEs as a function of Ly$\alpha$ luminosity. The
trends of stellar mass with luminosity and redshift are similar to
those already discussed for the halo mass: the trend of stellar mass
with luminosity steepens with increasing redshift, and the mass at a
given luminosity decreases with increasing redshift. The reasons
for the rather flat trend of stellar mass with Ly$\alpha$ luminosity
at the lower redshifts are the same as those already given for the
halo mass.  For a luminosity $L_{Ly\alpha} = 10^{42.5}h^{-2}\ergs$,
the median stellar mass is predicted to decrease from $\sim 10^9
h^{-1}\Msol$ at $z=3$ to $\sim 3\times 10^7 h^{-1}\Msol$ at $z=7$.

There have not yet been any observational estimates of the stellar
masses of LAEs, but the values predicted by our model are rather lower
than the stellar masses inferred observationally for some other
classes of high-redshift galaxies. \citet{Shapley01} estimated stellar
masses of Lyman-break galaxies at $z\sim 3$ from broad-band
photometry, and found a median stellar mass $\sim 10^{10} h^{-1}\Msol$
in a sample with median magnitude $R_{AB}\sim 24$ (similar results
were also found by \citealt{Papovich01}). In contrast, typical
observed LAEs at $z\sim 3$ with $L_{Ly\alpha} \sim 10^{42}
h^{-2}\ergs$ are predicted by our model to have stellar masses $\sim
10^9 h^{-1}\Msol$. The difference could be explained by a combination
of two effects: (a) The LAEs at $z\sim 3$ with $L_{Ly\alpha} \sim
10^{42} h^{-2}\ergs$ typically have fainter continuum magnitudes (by
1-2 mag) than the \citeauthor{Shapley01} LBGs. (b) The photometric
estimates of the stellar masses of LBGs depend strongly on the IMF
assumed, because the mass-to-light ratio of a stellar population is
very sensitive to the IMF; \citeauthor{Shapley01} assumed a Salpeter
IMF, but if instead they had assumed a top-heavy IMF as in starbursts
in our model, then they might have derived lower masses.

The issue of how photometric estimates of the stellar masses of
high-redshift galaxies depend on the assumed IMF is an important one,
but is also complicated, because these estimates involve fitting
multi-parameter models (varying age, star formation history,
metallicity and dust extinction) to multi-band photometric data, in
order to estimate the mass-to-light ratio of the stellar
population. Most such studies have simply assumed a Salpeter
IMF. \citet{Papovich01} considered the effects on photometric mass
estimates of varying the lower mass limit on the IMF, but did not
consider IMF slopes different from the solar neighbourhood value. We
plan to address this issue in more detail in a future paper.

\subsection{Metallicities}
\label{sec:metallicity}
The middle right panel of Fig.\ref{fig:props} shows the
metallicity of the cold gas in LAEs as a function of Ly$\alpha$
luminosity. We see that in most cases, the gas metallicity is
appreciable, $\sim 10^{-2}$ (i.e. comparable to solar), even at high
redshifts. This reflects the fact that galaxies are able to
self-enrich to metallicities $\sim \Zsol$ even when the mean
metallicity of all baryons in the universe is much lower than this. In
our model, the quiescent galaxies are predicted to show a well-defined
trend of metallicity increasing with luminosity (as already found for
$z=0$ galaxies by \citealt{Cole00}), which produces the decline in
metallicity at low luminosities seen in Fig.\ref{fig:props} for $z=3$
and $z=5$. However, the bursts show a more complicated behaviour, with
the median metallicity being flat or even non-monotonic with
luminosity, which is reflected in the behaviour seen in
Fig.\ref{fig:props} for $z=7$ and for higher luminosities at $z=3$ and
$z=5$.

\subsection{Star formation rates}
\label{sec:SFR}
The lower left panel of Fig.\ref{fig:props} shows the instantaneous
star formation rates in LAEs as a function of Ly$\alpha$
luminosity. The star formation rates include both the quiescent star
formation in galactic disks and the contribution of any ongoing
starbursts. If we look at either quiescent galaxies or bursts
separately, then we find a nearly linear relation $SFR \propto
L_{Ly\alpha}$, but with a proportionality constant which is $\sim 10$
times larger for the quiescent galaxies, because of the difference in
IMFs. For a galaxy which has been forming stars at a contant rate
for $10^8\yr$, we predict a relation $L_{Ly\alpha} = (1.1,12) \times
10^{42}\ergs f_{\rm esc} (SFR/\Msol\yr^{-1})$ for the Kennicutt and
$x=0$ IMFs respectively (for solar metallicity), and quiescent and
bursting galaxies in our model separately lie quite close to one or
other relation.  In Fig.\ref{fig:props} we see a shallower relation
than $SFR \propto L_{Ly\alpha}$ in some cases, which results from a
gradual transition from being dominated by bursts at high luminosities
to being dominated by quiescently star-forming galaxies at low
luminosities. For LAEs with luminosities $\sim 10^{42.5}h^{-2}\ergs$
(which are dominated by bursts), our model predicts SFRs $\sim 10
h^{-1}\Msol\yr^{-1}$. We note that comparing the SFRs predicted by our
model with published values estimated from observational data is not
straightforward, because different authors (a) assume different values
for the Ly$\alpha$ escape fraction (often taking $f_{\rm esc}=1$), and
(b) assume different IMFs (typically using a Salpeter IMF).

\subsection{UV continuum luminosities}
\label{sec:UV}
The lower right panel of Fig.\ref{fig:props} shows the UV
continuum luminosity $L_{\nu}(1500\AA)$ at a fixed rest-frame
wavelength of 1500\AA\ as a function of Ly$\alpha$ luminosity. This
plot contains similar information to Fig.\ref{fig:mags}, but now, for
convenience, in terms of luminosities, and at a fixed rest-frame
wavelength in the UV. The predicted relation between UV and Ly$\alpha$
luminosities is roughly linear. This is what one would expect if one
had a universal IMF and no dust attenuation, since both the UV and
Ly$\alpha$ luminosities are driven by recent star formation. For a
galaxy which has been forming stars at a contant rate for $10^8\yr$,
we predict an unattenuated relation $L_{\nu}(1500\AA) = (0.98,3.8)
\times 10^{28} \ergs\Hz^{-1} (SFR/\Msol\yr^{-1})$ for the Kennicutt
and $x=0$ IMFs respectively, for solar metallicity. However, dust
extinction is predicted to have a large effect on UV luminosities in
our model, reducing them by a factor $\sim 10$ in the brighter
objects. The average UV extinction in the models increases with
luminosity in both bursts and quiescent galaxies, and is also larger
in quiescent than bursting objects (at a given SFR). The unattenuated
UV/Ly$\alpha$ luminosity ratio is also predicted to decrease by a
factor 3 going from quiescent objects at low luminosity to bursts at
high luminosity as a result of the change in the IMF. These effects
all combine to leave a relation between UV and Ly$\alpha$ luminosities
which is shifted but still roughly linear.

The UV and Ly$\alpha$ luminosities are both used in estimating star
formation rates for observed high-redshift galaxies. However, both
suffer from the drawback that they are affected by large but uncertain
dust attenuation factors. In the model presented here, the dust
attenuation is larger by a factor $\sim 10$ for the Ly$\alpha$ than
for the UV luminosity, so by that criterion, the UV luminosity should
be a more reliable quantitative star formation indicator. However, we
caution that the Ly$\alpha$ attenuation factor $f_{\rm esc}$ which we
use was not obtained from a detailed radiative transfer calculation,
unlike the UV dust extinction.

\begin{figure}
\centering

{\epsfxsize=6truecm
\epsfbox{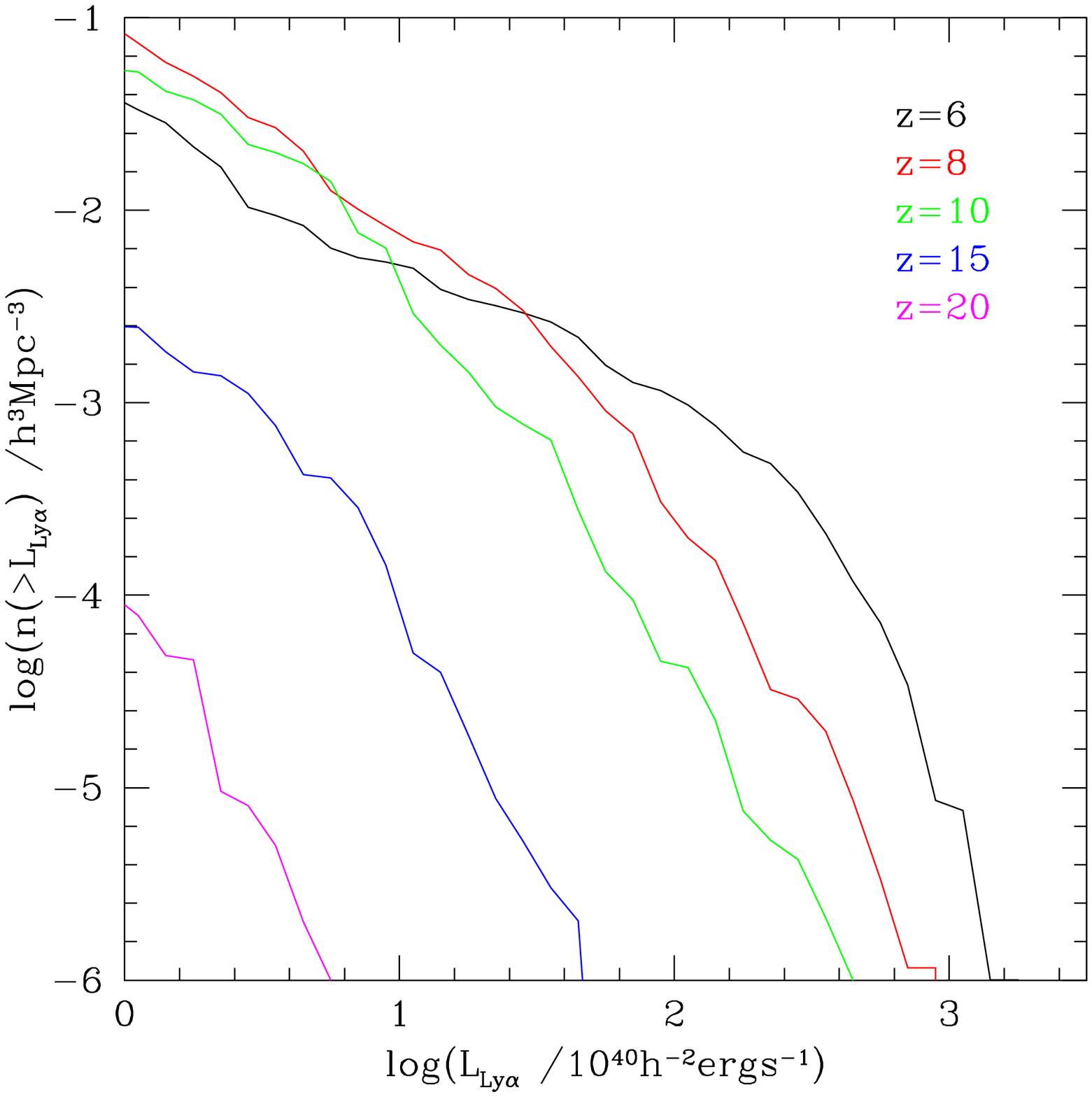}}
{\epsfxsize=6truecm
\epsfbox{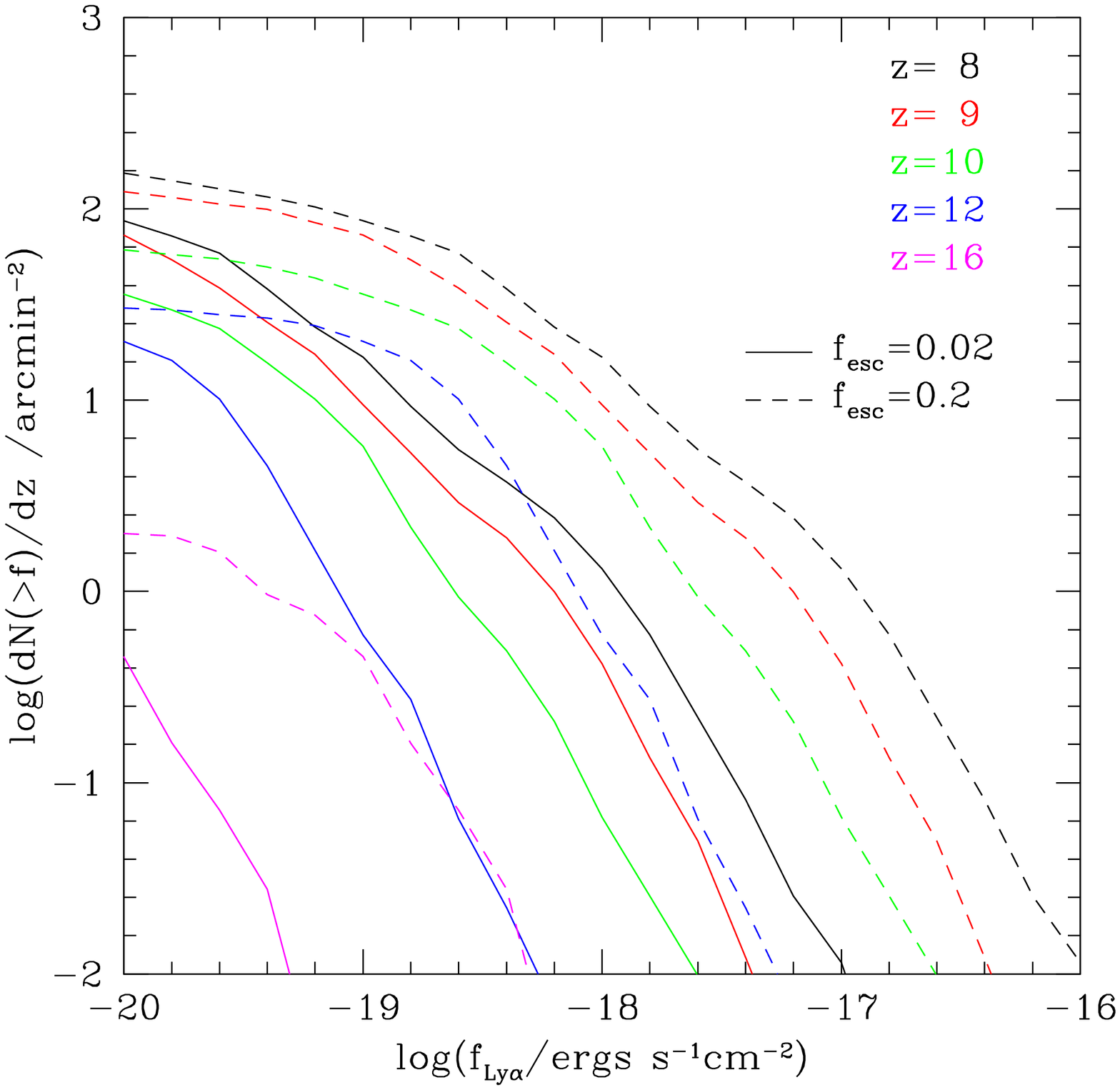}}
{\epsfxsize=6truecm
\epsfbox{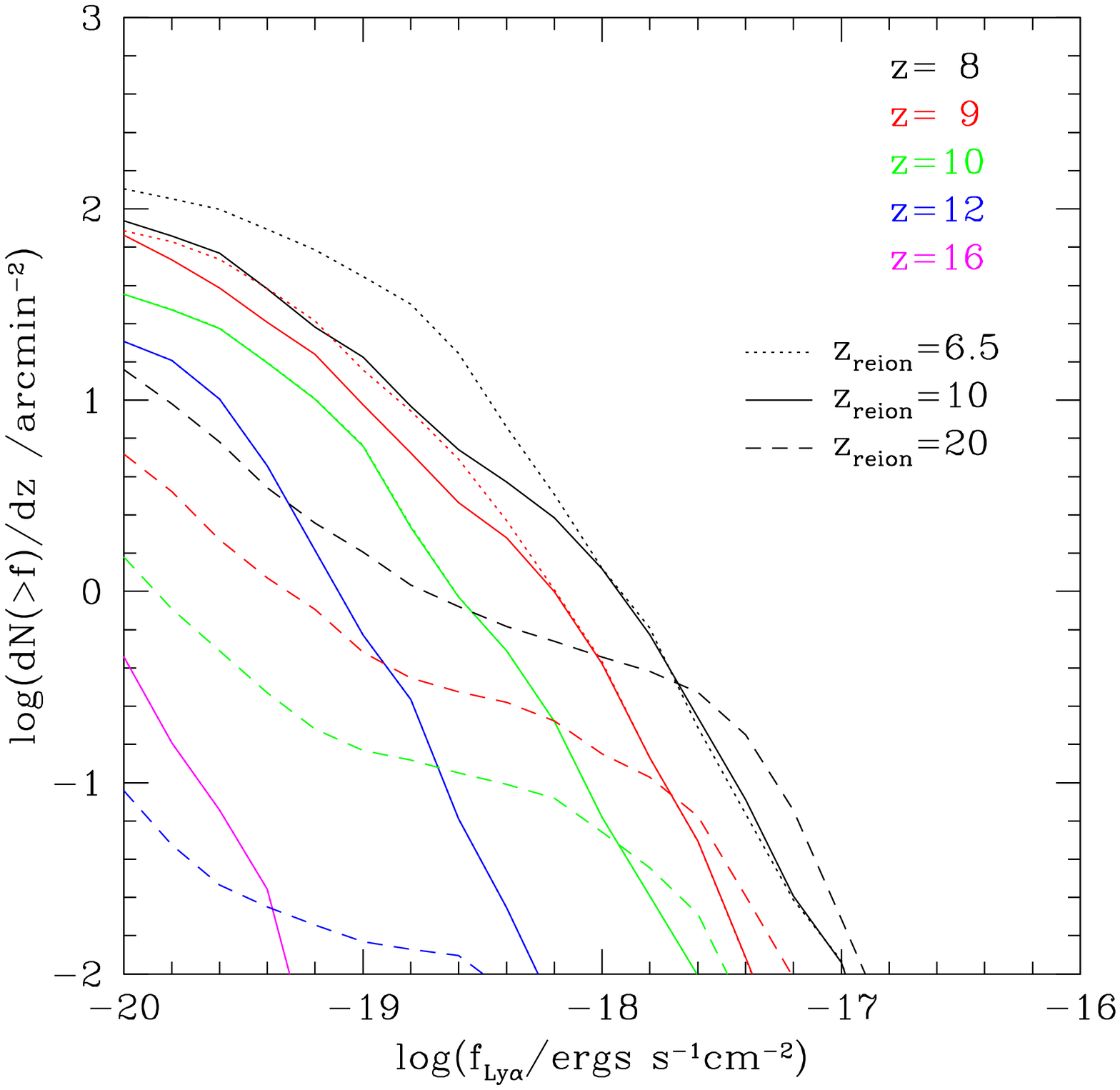}}

\caption{ Predictions for the number of Ly$\alpha$ emitters at very
high redshifts.  (a) The upper panel shows the evolution of the
cumulative Ly$\alpha$ luminosity function at $6 \leq z \leq 20$, for
our standard model, with reionization redshift $z_{\rm reion}=10$ and
Ly$\alpha$ escape fraction $f_{\rm esc}=0.02$. (b) The middle panel
shows the predicted number counts per unit redshift interval per unit
solid angle as a cumulative function of the Ly$\alpha$ flux, for
$z_{\rm reion}=10$. Results are shown for two different escape
fractions, our standard value $f_{\rm esc}=0.02$ (solid lines) and a
larger value $f_{\rm esc}=0.2$ (dashed lines). We show number counts
for selected redshifts falling within either the J,H or K atmospheric
transmission windows, with different redshifts in different
colours. (c) The lower panel shows the cumulative number counts as in
(b), for our standard $f_{\rm esc}=0.02$, and three different
reionization redshifts, $z_{\rm reion}=6.5$ (dotted lines), 10 (solid
lines) and 20 (dashed lines); where the dotted lines are not visible,
they coincide with the solid lines. The number counts do not include
attenuation by the IGM.}
\label{fig:LF_hiz}
\end{figure}

\section{The abundance of Ly$\alpha$ emitters at $z>7$}
\label{sec:high-z}
Paper~I presented predictions for the abundance of Ly$\alpha$ emitters
as a function of flux for the redshift range $2 \lsim z \lsim 7$ which
is accessible from observations at optical wavelengths. The highest
redshift at which LAEs have been found in surveys up to now is
$z=6.6$. At even higher redshifts, the Ly$\alpha$ line moves into the
near-IR as seen from the Earth. Therefore, searching for LAEs at
$z\gsim 7$ requires observing in the near-IR, which is technically
challenging. Several such searches are underway
\citep[e.g.][]{Willis05a,Stark05}, and others will start in the near
future \citep[e.g.][]{DAZLE}.
Therefore in this section we present some predictions for the number
of LAEs which should be found in near-IR searches, at redshifts $7
\lsim z \lsim 20$. \citet{Barton04} have previously made predictions
for the number of LAEs at $z \approx 8$ based on a numerical
simulation, but assumed a 100\% escape fraction for Ly$\alpha$ photons
(i.e. $f_{\rm esc} =1$). \citet{Thommes05} have also made predictions
for $z>7$, but for a phenomenological model not based on CDM.

The upper panel of Fig.\ref{fig:LF_hiz} shows what our standard model,
with reionization redshift $z_{\rm reion}=10$ and Ly$\alpha$ escape
fraction $f_{\rm esc}=0.02$, predicts for the evolution of the
luminosity function of LAEs at $6 \leq z \leq 20$. We see that the
luminosity function declines significantly at the bright end from
$z\sim 6$ to $z\sim 10$, and then declines with redshift very rapidly
at all luminosities at $z\gsim 10$. This decline is driven by the
reduction in star formation with increasing redshift, which results
from the build-up of cosmic structure over time.

The middle and lower panels of Fig.\ref{fig:LF_hiz} show the predicted
number counts per unit solid angle per unit redshift as a cumulative
function of the Ly$\alpha$ flux. These predictions do not include
the attenuation of the Ly$\alpha$ flux by neutral gas in the
intervening IGM, as we discuss below. We show predictions for
redshifts $z=8,9,10,12$ and $16$, chosen such that the Ly$\alpha$ line
falls within either the J, H or K atmospheric window. (The J, H and K
atmospheric windows cover the wavelength ranges 1.08-1.35, 1.51-1.80
and 1.97-2.38 $\mu$m respectively, corresponding to Ly$\alpha$
redshift ranges $z=7.9-10.1$, $11.4-13.8$ and $15.2-18.6$.) 
Ground-based searches in the near-IR are likely to concentrate on
these atmospheric windows, because the atmospheric opacity at other
near-IR wavelengths is extremely high. We have already shown
predicted angular sizes for these galaxies in
Fig.\ref{fig:sizes}(b).

The middle panel of Fig.\ref{fig:LF_hiz} shows how the predicted
number counts depend on the assumed Ly$\alpha$ escape fraction $f_{\rm
esc}$. We show results for our standard value of the reionization
redshift $z_{\rm reion}=10$, but for two values of $f_{\rm esc}$, 0.02
(our standard value) and 0.2. We recall that the value of $f_{\rm
esc}$ was originally chosen in order to match the observed counts at
$z\sim 3$, and turns out to provide a good match to observations over
the whole range $3 < z < 6.6$. At higher redshifts, no empirical
calibration of $f_{\rm esc}$ is available. Since we do not have a
detailed physical model for $f_{\rm esc}$, we cannot exclude the
possibility that the value at very high redshifts is different from the
value at lower redshifts. As might be expected, the number counts at a
given flux are quite sensitive to the value of $f_{\rm
  esc}$. Predicted counts for other values of $f_{\rm esc}$ than those
shown in Fig.\ref{fig:LF_hiz} can easily be obtained by scaling the
curves in the horizontal direction with $f_{\rm esc}$, since the flux
from each galaxy is proportional to $f_{\rm esc}$.

The lower panel of Fig.\ref{fig:LF_hiz} shows how the predicted number
counts depend on the assumed reionization redshift $z_{\rm reion}$. As
was discussed in Section~\ref{sec:GALFORM}, observations of
Gunn-Peterson absorption troughs in QSO spectra suggest that the IGM
became fully reionized at $z \sim 6$ \citep{Becker01}, while the WMAP
measurement of the polarization of the microwave background implies
that the IGM has been mostly reionized since $z \sim 20$
\citep{Kogut03}. We therefore show predictions for $z_{\rm reion} =
6.5$ and 20, in addition to our standard value $z_{\rm reion} =
10$. In all three cases we assume $f_{\rm esc}=0.02$. In our model,
reionization is assumed to affect galaxy formation in the following
way: the IGM is assumed to be instantaneously reionized and reheated
at $z=z_{\rm reion}$, and at $z<z_{\rm reion}$, the thermal pressure
of the IGM is assumed to prevent gas collapsing in all dark halos with
circular velocities $V_c < 60 \kms$. This simple behaviour is an
approximation to what was found in more detailed calculations by
\cite{Benson02}. The dependence of the number counts on $z_{\rm
reion}$ shown in Fig.\ref{fig:LF_hiz} results entirely from this
effect of reionization on galaxy formation, since we ignore the IGM
opacity here. For $z>z_{\rm reion}$, all models look identical to the
case in which the IGM never reionized, but for $z<z_{\rm reion}$, the
number counts are suppressed relative to the no-reionization case. We
see that the predicted counts for the redshift ranges and fluxes shown
in Fig.\ref{fig:LF_hiz} differ only slightly for $z_{\rm reion} = 6.5$
and $z_{\rm reion} = 10$, but for $z_{\rm reion} = 20$, the predicted
counts are much lower, except for the bright counts at $z \sim 8 -
10$.

However, as noted earlier, there is an important caveat to our
results: our model includes the attenuation of the stellar continuum
light from galaxies due to Ly$\alpha$ scattering and Lyc absorption by
atomic hydrogen in the intervening IGM, but we do not include any
attenuation of the flux in the Ly$\alpha$ line due to Ly$\alpha$
scattering in the IGM. At $z > z_{\rm reion}$, when the IGM is
completely neutral, this attenuation of the Ly$\alpha$ line flux could
potentially be very large, which would greatly decrease the number
counts at a given flux. IGM attenuation will therefore produce a trend
in the number counts with $z_{\rm reion}$ in the opposite sense to the
feedback of reionization on galaxy formation: at $z<z_{\rm reion}$,
the feedback effect tends to suppress the counts, but the IGM
attenuation will be removed, which will increase the counts relative
to a model in which the IGM is still neutral at that redshift.

The amount of attenuation of the Ly$\alpha$ line by the IGM before
reionization is theoretically very uncertain. \citet{Miralda98} showed
that for a galaxy at high redshift embedded in a neutral IGM moving
with the Hubble flow, emitting a Ly$\alpha$ line centred at its
rest-frame wavelength in the frame of the galaxy, scattering by atomic
hydrogen in the IGM would suppress the blue wing of the Ly$\alpha$
line completely (reducing the line flux by a factor 2), and also
partly suppress the red wing due to the Ly$\alpha$ damping wings
(reducing the line flux even more). However, \citet{Madau00} and
\citet{Haiman02} showed that this attenuation of the line flux could
be greatly reduced due to the galaxy photo-ionizing the IGM around
it. A more detailed theoretical analysis of the attenuation has been
made by \citet{Santos04a}, who includes the following effects: (i) the
intrinsic width of the Ly$\alpha$ line emitted by the galaxy, and the
fact it may be redshifted in the galaxy rest-frame due to scattering
in a galactic wind; (ii) the non-uniform density profile of the IGM
around the galaxy and the departure of the velocity field from the
Hubble flow, due to cosmological infall onto the galaxy; (iii)
collisional ionization of the gas within the galaxy halo,
photo-ionization of the surrounding IGM by the stellar population, and
clearing of bubbles in the IGM by galactic
winds. \citeauthor{Santos04a} finds that a very wide range of
attenuation factors is possible in plausible models, but that if the
Ly$\alpha$ emission is redshifted in the rest-frame of the galaxy (as
is observed to be the case in Lyman-break galaxies at $z\sim 3$), the
amount of attenuation is greatly reduced. The effects on the
attenuation of Ly$\alpha$ of clustering of galaxies and clumping of
the IGM have been investigated by \citet{Furlanetto04}
\citet{Gnedin04}, and \citet{Wyithe05}; 
they find that these effects can also significantly reduce the amount
of attenuation. In summary, for LAEs at $z > z_{\rm reion}$, the
Ly$\alpha$ flux is likely to be significantly attenuated by the IGM,
but the amount of attenuation is currently uncertain. We plan to
investigate this in more detail in a future paper.

An unsuccessful search for LAEs at $z=8.8$ has been carried out by
\citet{Willis05a}, who place an upper limit $n \lsim 3\times 10^{-3}
h^3 \Mpc^{-3}$ on sources with $L_{Ly\alpha} > 10^{43}
h^{-2}\ergs$. This limit on the number density is more than 4 orders
of magnitude higher than the prediction for our standard model shown
in Fig.\ref{fig:LF_hiz}(a). \citet{Willis05b} and \citet{Stark05} are
carrying out surveys for $z \sim 9$ LAEs which will probe to lower
luminosities by using gravitational lensing by foreground clusters. 
In the near future, the DAzLE instrument \citep{DAZLE} on the VLT will
begin searches for LAEs at $z>7$. It is planned to search for
Ly$\alpha$ at $z \approx 7.7$ in the first phase, and $z \approx 8.7$
in the second phase, in each case over a redshift window $\Delta z
\approx 0.01$; ultimately it may be possible to reach $z \approx 14$
with the instrument. DAzLE will cover an area of $\Delta\Omega = 47\,
{\rm arcmin}^2$ in a single exposure, and is projected to reach a
$5\sigma$ flux limit of $f = 2 \times 10^{-18} \ergcms$ in a 10 hour
integration. We can use our model to predict how many objects DAzLE
should see in a single 10 hour exposure in each of the redshift
ranges. For our standard model with $z_{\rm reion}=10$ and $f_{\rm
esc}=0.02$, we predict the number of sources per unit redshift per
unit solid angle above flux $f = 2 \times 10^{-18} \ergcms$ to be
$d^2N(>f)/d\Omega dz = (0.58,0.13)\, {\rm arcmin}^{-2}$ at
$z=(7.7,8.7)$ respectively, which for the specified $\Delta z$ and
$\Delta\Omega$ corresponds to an average of 0.3 and 0.06 objects per
field at $z=7.7$ and $z=8.7$ respectively. Therefore we expect that
DAzLE will need to observe many separate fields to find a significant
number of high-z Ly$\alpha$ emitters.

\section{Conclusions}
\label{sec:conc}
In this paper, we have used a detailed semi-analytical model of galaxy
formation based on the $\Lambda$CDM cosmology to predict the
properties of star-forming Ly$\alpha$-emitting galaxies over the
redshift range $0 \leq z \leq 20$. All except one of the parameters of
the model were chosen without reference to the observed properties of
Ly$\alpha$ emitters, having instead been chosen in previous work
\citep{Cole00, Baugh05} to match properties such as the UV, optical
and IR luminosities, sizes, morphological types, gas fractions and
metallicities of galaxies at low and high redshift. As shown in
\citet{Baugh05}, our current model, which incorporates a top-heavy IMF
for stars formed in bursts triggered by galaxy mergers, provides a
good match to the optical and far-IR luminosity functions in the local
universe, the far-UV luminosity function of Lyman-break galaxies at
$z\sim 3$, and the number counts and redshifts of sub-mm galaxies. Our
assumption of a top-heavy IMF in bursts receives  support from
studies of the metallicities of elliptical galaxies and intracluster
gas \citep{Nagashima05a,Nagashima05b}. The one free parameter we had
in the comparison of our model with observational data on Ly$\alpha$
emitters (LAEs) was the fraction $f_{\rm esc}$ of Ly$\alpha$ photons
which escape from a galaxy. For simplicity, we have assumed that
$f_{\rm esc}$ is a constant, irrespective of other galaxy
properties. In our previous paper on LAEs \citep{LeD05}, we found that
$f_{\rm esc}=0.02$ reproduced the abundance of faint LAEs at $z\sim
3$, and we used the same value in the present paper.

In \citet{LeD05}, we presented model predictions for the number counts
of LAEs as a function of Ly$\alpha$ flux for the redshift range $2\leq
z \leq 6$, but made only a very limited comparison with observational
data, comparing only with the total counts at the limiting fluxes for
different surveys. In this paper, we have made a much more detailed
comparison of the model with observations, comparing predicted and
observed Ly$\alpha$ luminosity functions over the range $3 \leq z \leq
7$. The most important result of this paper is that, with our very
simple assumption of a constant Ly$\alpha$ escape fraction, the model
reproduces the observed luminosity functions, both in shape and in the
evolution with redshift.

We have also compared the predictions of our model with other observed
properties of LAEs. We have made a  comparison of predicted and
observed Ly$\alpha$ equivalent widths (EWs). We find that the typical
predicted EWs are similar to those found in observational surveys, and
that the predicted distribution of EWs at a given Ly$\alpha$ flux is
very broad once we include the effect of dust extinction, with a peak
at 0 and a tail to large values. If we select galaxies in the model
according to their continuum magnitudes rather than their Ly$\alpha$
fluxes, then we predict a distribution of EWs which is similar to what
has been found observationally for Lyman-break galaxies at $z \sim 3$,
when we restrict the comparison to galaxies where Ly$\alpha$ is seen
in emission. We have compared predicted and observed broad-band
magnitudes (corresponding to rest-frame UV luminosities) for galaxies
selected by their Ly$\alpha$ fluxes, and find mostly good
agreement. We have also compared the predicted sizes of LAEs with the
limited existing observational data, and find reasonable consistency
for the stellar half-light radii.

We have also used our model to try to better understand the nature of
the objects selected in Ly$\alpha$ emission-line surveys, and how they
relate to other classes of high-redshift galaxies. We have made
predictions for the dark halo and stellar masses, the star formation rates
and the gas metallicities for LAEs at different redshifts, properties
which are physically fundamental, but for which we do not yet have
direct observational measurements. 
The predicted halo masses imply values of the clustering bias which
seem quite consistent with existing measurements of the large-scale
clustering of LAEs. Better observational characterization of the
clustering of LAEs at different redshifts would provide a very
important test of our model.

Finally, we have presented predictions for how many Ly$\alpha$
emitters should be seen at $z>7$, a redshift range for which no
LAEs have yet been found, but which is now opening up for
observational study, thanks to advances in near-IR
instrumentation. Detection of LAEs in this redshift range would be
very exciting both for probing the early stages of galaxy formation
and the epoch of reionization. A problem for making predictions for
LAEs at $z>7$ is that the redshift at which the IGM reionized is
uncertain, being observationally constrained to be in the range $6.5
\lsim z_{\rm reion} \lsim 20$. This affects predictions for galaxies
seen at $z> z_{\rm reion}$, since the Ly$\alpha$ flux is expected to
be significantly attenuated by propagation through a neutral IGM, but
by an uncertain amount which depends on many factors. We have made
detailed predictions for how many objects could be seen using the
DAzLE instrument, which begins operation soon, at $z \approx 7.7$ and
$z \approx 8.7$. We find that, even if the attenuation by the IGM is
modest at these redshifts, then finding LAEs with DAzLE will require
observation of a large number of fields.

The most important theoretical limitations of our present work are
that it does not incorporate a detailed physical model for the escape
of Ly$\alpha$ photons from galaxies, and that we do not include a
treatment of the attenuation of Ly$\alpha$ fluxes by the IGM prior to
reionization. We plan to address these issues in future papers.

\section*{Acknowledgements} 
We thank Kim-Vy Tran and Simon Lilly for providing us with their
luminosity function compilations in convenient form.  MLeD was
supported by the Royal Society through an Incoming Short Visit
award. CMB acknowledges the receipt of a Royal Society University
Research Fellowship. This work was also supported by the PPARC rolling
grant for extragalactic astronomy and cosmology at Durham. We thank
the anonymous referee for a constructive report.

\bsp
\end{document}